\documentclass[twocolumn,prl,twocolumn]{revtex4-1}
\usepackage[latin9]{inputenc}
\usepackage{color}
\usepackage{amsmath}
\usepackage{amssymb}
\usepackage{graphicx}
\usepackage[unicode=true,
 bookmarks=true,bookmarksnumbered=false,bookmarksopen=false,
 breaklinks=false,pdfborder={0 0 1},backref=false,colorlinks=true]
 {hyperref}
\hypersetup{
 linkcolor=magenta, urlcolor=blue, citecolor=blue, pdfstartview={FitH}, hyperfootnotes=false, unicode=true}

\makeatletter

\usepackage{amsfonts}\usepackage{tabularx}\usepackage{dcolumn}\usepackage{bm}\usepackage{graphicx}\usepackage{epstopdf}
\usepackage{times}
\hypersetup{urlcolor=blue}

\makeatother

\begin{document}

\title{Machine Learning Bell Nonlocality in Quantum Many-body Systems}

\author{Dong-Ling Deng}

\affiliation{Condensed Matter Theory Center and Joint Quantum Institute, Department
of Physics, University of Maryland, College Park, MD 20742-4111, USA}
\begin{abstract}
Machine learning, the core of artificial intelligence and big data
science, is one of today\textquoteright s most rapidly growing interdisciplinary
fields. Recently, its tools and techniques have been adopted to tackle
intricate quantum many-body problems. In this work, we introduce machine
learning techniques to the detection of quantum nonlocality in many-body
systems, with a focus on the restricted-Boltzmann-machine (RBM) architecture.
Using reinforcement learning, we demonstrate that RBM is capable of
finding the maximum quantum violations of multipartite Bell inequalities
with given measurement settings. Our results build a novel bridge
between computer-science-based machine learning and quantum many-body
nonlocality, which will benefit future studies in both areas.
\end{abstract}
\maketitle
\begin{figure}
\includegraphics[width=0.46\textwidth]{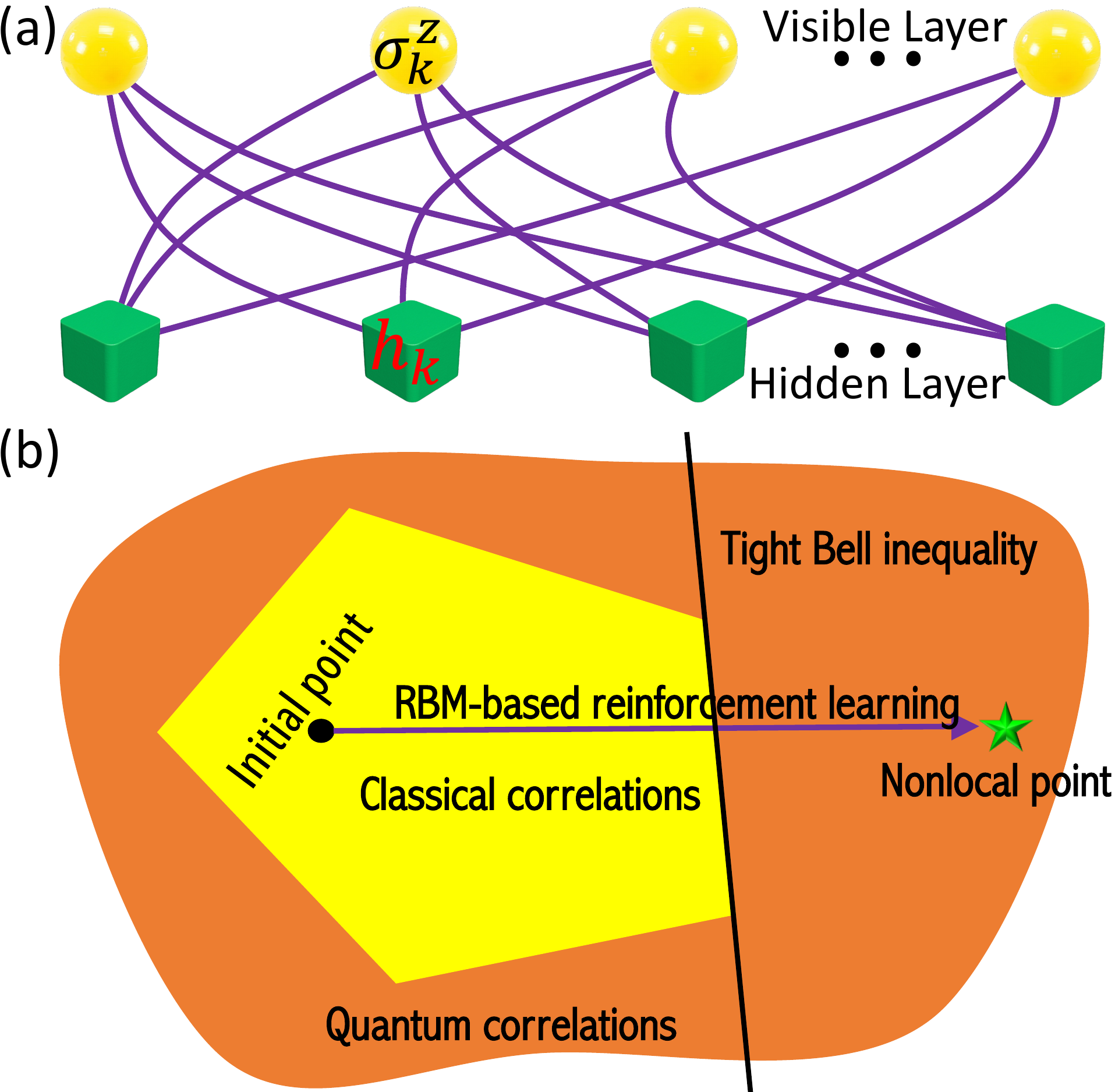}

\caption{(a) A sketch of the restricted-Boltzmann-machine (RBM) representation
of quantum many-body states. For each spin configuration $\Xi=(\sigma_{1}^{z},\sigma_{2}^{z},\cdots,\sigma_{N}^{z})$,
the artificial neural network outputs its corresponding coefficient
$\Phi(\Xi$). (b) A pictorial illustration of the essential idea of
machine learning Bell nonlocality in quantum many-body systems. The
set of all classical correlations forms a high-dimensional polytope
(yellow region), which is a subset of the quantum-correlation set
that consists of all possible correlations allowed by quantum mechanics.
The black line represents a tight Bell inequality (facet of the polytope).
We start with a random RBM, which typically shows only classical correlations
(in the sense that it does not violate a given Bell inequality). We
then optimize its internal parameters, through reinforcement learning,
so as to violate the Bell inequality maximally. \label{fig:MLBNonL}}
\end{figure}

Nonlocality is one of the most fascinating and enigmatic features
of quantum mechanics that denies any local realistic description of
our world \cite{Bell1964OnEPR,Brunner2014Bell}. It represents the
most profound departure of quantum from classical physics and has
been experimentally confirmed in a number of systems through violations
of Bell inequalities \cite{Freedman1972Experimental,Aspect1982Experimental,Weihs1998Violation,Rowe2001Experimental,Giustina2013Bell,Christensen2013Detection,Eibl2003Experimental,Zhao2003Experimental,Lanyon2014Experimental,Hofmann2012Heralded,Pfaff2013Demonstration,Ansmann2009Violation,Hensen2015Loophole,Giustina2015Significant,Shalm2015Strong,Zu2013Experimental}.
In addition to its fundamental interest, in practice nonlocality is
the key resource for device-independent quantum technologies, such
as secure key distribution \cite{Acin2007Device,Pironio2013Security,Vazirani2014Fully}
or certifiable random number generators \cite{Colbeck2007Quantum,Pironio2010Random,Deng2013Fault-tolerant,Herrero2017Quantum,Miller2014Robust}.
Thus, characterizing and detecting nonlocality is one of the central
problems in both quantum information theory and experiment. Here,
we introduce machine learning, a branch of computer science \cite{Michalski2013Machine,Jordan2015Machine,Lecun2015Deep},
to the detection of quantum nonlocality (see Fig. \ref{fig:MLBNonL}
for a pictorial illustration).

For quantum many-body systems, whereas entanglement has been extensively
studied \cite{Amico2008Entanglement}, nonlocality remains rarely
explored. Mathematically, it has been proved that the complete characterization
of classical correlations for a generic many-body system is an NP-hard
problem \cite{Babai1991Non}. Nevertheless, an incomplete list of
multipartite Bell inequalities with high-order correlation functions
has indeed been discovered for a long time \cite{Brunner2014Bell}.
More recently, Bell inequalities with only two-body correlators were
constructed \cite{Tura2014Detecting,Tura2015Nonlocality,Tura2017Energy,Wang2017Entanglement,Wagner2017Bell}
and multipartite nonlocality has been demonstrated experimentally
in a Bose-Einstein condensate by violating one of these inequalities
\cite{Schmied2016Bell}. This sparks a new wave of interest in the
study of nonlocality in many-body systems. 

A particular question of both theoretical and experimental relevance
is that for a given multipartite Bell inequality, how to obtain its
quantum violation in a numerical simulation? To tackle this problem,
one has to face at least two challenges. First of all, the Hilbert
space of a quantum many-body system grows exponentially with the system
size and a complete description of its state requires an exponential
amount of information in general, rendering the computation of the
quantum expectation value corresponding to the inequality a formidably
demanding task. Second, the measurement settings for each party involved
in a Bell experiment is arbitrary in principle, making the problem
even more complicated. In fact, it has been shown that the computation
of the maximum violation of a multipartite Bell inequality is an NP-problem
\cite{Batle2016Computing}. In this paper, we will not attempt to
solve this problem completely, which is implausible due to the NP
complexity. Instead, we study a simplified scenario where the given
multipartite Bell inequality only involves a polynomial number of
correlation functions and the measurement settings for each party
are restricted (due to experimental requirements, for instance) and
preassigned. We show that machine learning may provide an unprecedented
perspective for solving this simplified, but still sufficiently intricate,
quantum many-body problem. Within physics, applications of machine-learning
techniques have recently been invoked in various contexts \cite{Carleo2016Solving,Arsenault2015Machine,Zhang2016Triangular,Carrasquilla2017Machine,van2017Learning,Deng2016Exact,Wang2016Discovering,Broecker2017Machine,Chng2017Machine,Zhang2017Machine,Wetzel2017Unsupervised,Hu2017Discovering,Yoshioka2017Learning,Torlai2016Learning,Aoki2016restricted,You2017Machine,Torlai2017Many,Pasquato2016Detecting,Hezaveh2017Fast,Rahul2013Application,Abbott2016Observation,Kalinin2015Big,Schoenholz2016Structural,Liu2017Self,Huang2017Accelerated,Torlai2017Neural,Gao2017Efficient,Chen2017Equivalence,Huang2017Neural,Schindler2017Probing,Cai2017Approximating,Broecker2017Quantum,Nomura2017Restricted,Biamonte2017Quantum},
such as black hole detection \cite{Pasquato2016Detecting}, gravitational
lenses \cite{Hezaveh2017Fast} and wave analysis \cite{Rahul2013Application,Abbott2016Observation},
material design \cite{Kalinin2015Big}, glassy dynamics \cite{Schoenholz2016Structural},
Monte Carlo simulation \cite{Liu2017Self,Huang2017Accelerated}, topological
codes \cite{Torlai2017Neural}, quantum machine learning \cite{Biamonte2017Quantum},
and topological phases and phase transitions \cite{Zhang2016Triangular,Carrasquilla2017Machine,van2017Learning,Wang2016Discovering,Broecker2017Machine,Chng2017Machine,Zhang2017Machine,Deng2016Exact,Wetzel2017Unsupervised,Hu2017Discovering},
etc. Here, we apply machine learning to the detection of quantum many-body
nonlocality, focusing on one of the simplest stochastic neural networks
for unsupervised learning\textemdash the restricted Boltzmann machine
(RBM) \cite{Hinton2006Reducing,Salakhutdinov2007Restricted,Larochelle2008Classification}
as an example. We demonstrate, through three concrete examples, that
RBM-based reinforcement learning is capable of finding the maximum
quantum violations of multipartite Bell inequalities with given measurement
settings. Our method works for generic Bell inequalities that involve
a polynomial number (in system size) of correlation functions, independent
of dimensionality, the order of the correlations, or whether the correlation
functions are short-range or not. Our results showcase the exceptional
power of machine learning in the detection of quantum nonlocality
for many-body systems, thus would provide a valuable guide for both
theory and experiment.

To begin with, let us first briefly introduce the RBM representation
of quantum states \cite{Carleo2016Solving} and the general recipe
for machine learning Bell nonlocality. We consider a quantum system
with $N$ spin-$\frac{1}{2}$ particles (qubits) $\Xi=(\sigma_{1},\sigma_{2},\cdots\sigma_{N})$
and use a RBM neural network to describe its many-body wave-function
$\Phi(\Xi)$. A RBM consists of two layers: one called visible layer
with $N$ nodes (visible neurons), corresponding to the physical spins;
the other called hidden layer with $M$ auxiliary nodes (hidden neurons).
The hidden neurons are coupled to the visible ones, but there is no
coupling among neurons in the same layer, as schematically illustrated
in Fig. \ref{fig:MLBNonL}(a). By tracing out the hidden neurons,
we obtain a RBM representation of a quantum many-body state \cite{Carleo2016Solving}:
\begin{eqnarray}
\Phi_{M}(\Xi,\Omega) & = & \sum_{\{h_{k}\}}e^{\sum_{k}a_{k}\sigma_{k}^{z}+\sum_{k'}b_{k'}h_{k'}+\sum_{kk'}W_{k'k}h_{k'}\sigma_{k}^{z}},\quad\label{eq:RBMRepresentation}
\end{eqnarray}
where $\Omega\equiv(a,b,W)$ are internal parameters that fully specify
the RBM neural network and $\{h_{k}\}=\{-1,1\}^{M}$ denotes the possible
hidden neuron configurations. We mention that any quantum state can
be approximated to arbitrary accuracy by the above RBM representation,
as long as the number of hidden neurons is large enough \cite{Kolmogorov1963Representation,Le2008Representational,Hornik1991Approximation}.
It is shown in Ref. \cite{Deng2016Exact} that RBM can represent topological
states, either symmetry protected or with intrinsic topological order,
in an exact and efficient fashion, and the entanglement properties
of RBM states are extensively studied in Ref \cite{Deng2017Quantum}. 

We consider a standard Bell experiment in which $N$ parties each
can freely choose to perform one of $K$ possible measurements $\mathcal{M}_{k}^{(i)}$
($i=1,\cdots,N\;\text{and }k=0,\cdots,K-1)$ with binary outcomes
$\pm1$. We describe the observed correlations by using a collection
of expectation values of correlators $\langle\mathcal{M}_{k_{1}}^{(i_{1})}\cdots\mathcal{M}_{k_{\alpha}}^{(i_{\alpha})}\rangle$
and we say that the correlations are classical when they can be simulated
with only shared classical information between parties (or in other
words, can be described by a local hidden variable theory \cite{Einstein1935Can}).
Classical correlations form a high-dimensional (exponential in $N$)
polytope $\mathbb{P}$, which is a bounded convex set with a finite
number of extreme points. Each facet of $\mathbb{P}$ corresponds
to a tight Bell inequality and correlations that fall outside of $\mathbb{P}$
will violate a Bell inequality and thus manifest nonlocality. We write
the Bell inequalities in a generic form: $\mathcal{I}\geq\mathcal{B}^{(c)}$,
where $\mathcal{I}$ is a function of the expectation values of the
correlators and $\mathcal{B}^{(c)}$ is the classical bound. Within
this framework, our general recipe for machine learning of nonlocality
through violation of a given Bell inequality is as follows: we begin
with a random RBM state, whose observed correlations may or may not
fall inside $\mathbb{P}$, but typically do not violate the given
inequality; we then use a reinforcement learning scheme recently introduced
by Carleo and Troyer \cite{Carleo2016Solving} to iteratively optimize
the internal parameters, such that the minimal expectation value of
$\mathcal{I}$ within quantum mechanics will be achieved. If the minimal
value is smaller than $\mathcal{B}^{(c)}$, the Bell inequality is
maximally violated with a given measurement setting and nonlocality
is detected. A pictorial illustration of the classical polytope, a
tight Bell inequality, and the essential idea of machine learning
Bell nonlocality is shown in Fig. \ref{fig:MLBNonL}(b). 

One may also choose another measurement setting and run the same process
to obtain the maximal violation for this setting. In order to obtain
the maximal violation of the Bell inequality for all measurement settings,
one can just scan all possible settings and do the same process repeatedly.
We mention that an alternative and more efficient way is to regard
all the parameters that specify the measurements as variational parameters
as well (on an equal footing as the RBM parameters $\Omega$) and
optimize them together with the RBM parameters using a similar reinforcement
learning procedure. But this is more technically involved. Here, we
will only focus on the former cases with fixed measurement settings
(parameters for measurements are preassigned) for simplicity and leave
the later approach for future studies. 

To show more precisely how this RBM-based reinforcement learning protocol
works, we give three concrete examples. The first one concerns Bell
inequalities with only short-range two-body correlators in one dimension
(1D). This is a case where traditional methods, such as density-matrix
renormalization group (DMRG), also work remarkably well \cite{White1992Density,Schollwock2011Density,Schollwock2005TheDMRG}.
We compare our RBM results with that from exact diagonalization (ED)
for small system size $N$ and DMRG for larger $N$, and find that
they agree excellently. This validates the effectiveness of our RBM
approach. The second and third examples are about Bell inequalities
with, either all-to-all but two-body or multipartite, correlators.
These examples are beyond the capacity of the DMRG or ED methods for
large system sizes and show a striking advantage of RBMs in detecting
many-body nonlocality. 

\begin{figure*}
\includegraphics[width=0.98\textwidth]{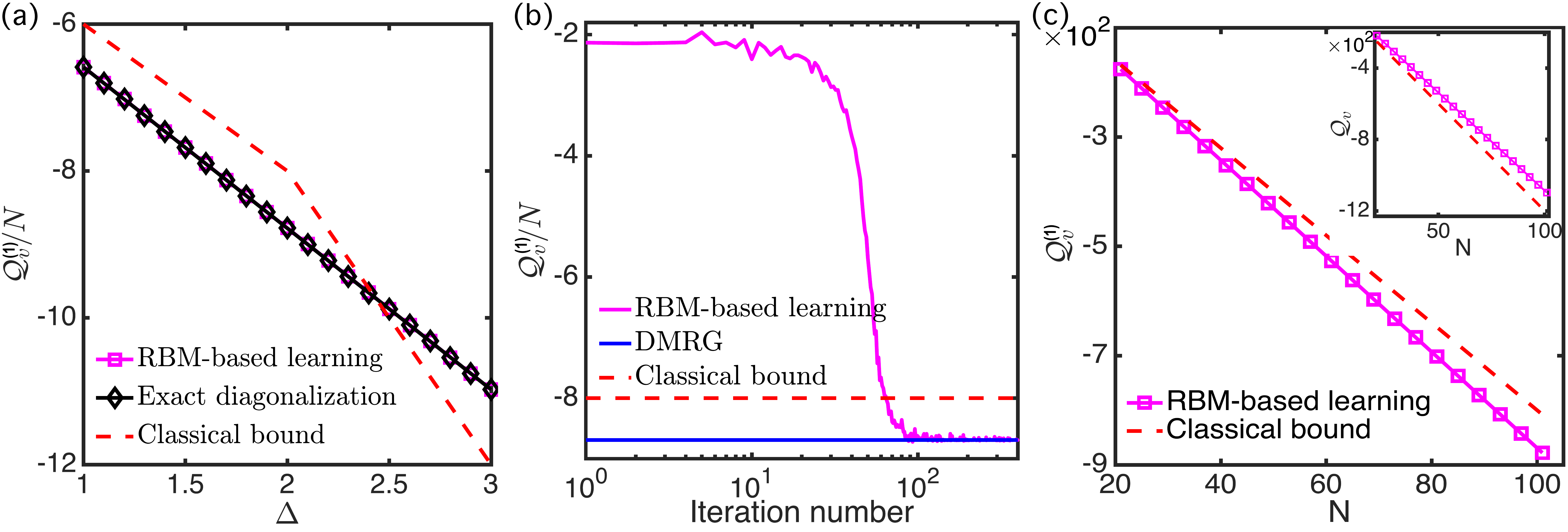}

\caption{RBM-based reinforcement learning of many-body Bell nonlocality through
quantum violations of the Ineq. (\ref{BIsr2bc}). The red dashed lines
represent the classical bounds $\mathcal{B}_{1}^{(c)}$, the regions
below which show quantum nonlocality and thus are not attainable by
any local hidden variable models. We denote the quantum expectation
value corresponding to $\mathcal{I}_{1}$ as $\mathcal{Q}_{v}^{(1)}$,
and without loss of generality, we have fixed $\delta=0.9$ throughout
this figure for simplicity. (a) A comparison between results from
RBM and exact diagonalization (ED) for a small system size $N=20$.
The results match each other very well. (b) The obtained quantum expectation
value as a function of the iteration number of the learning process,
for a larger system size $N=100$. For this particular learning process,
$\mathcal{Q}_{v}^{(1)}$ begins to cross the classical bound $\mathcal{B}_{1}^{(c)}$
after $65$ iterations, and all the RBM states thereafter violate
Ineq. (\ref{BIsr2bc}) and thus show many-body nonlocality. As the
iteration number increases, $\mathcal{Q}_{v}^{(1)}$ converges quickly
to the value computed from DMRG \cite{BvioSup}. (c) RBM learned $\mathcal{Q}_{v}^{(1)}$
as a function of $N$ for $\Delta=2$. The inset shows the result
for $\Delta=3$, where no quantum violation is observed.\label{fig:Reinforcement-learning-XXZ}}
\end{figure*}

\textit{Bell inequalities with short-range two-body correlators.}\textemdash Now,
let us consider a 1D system with $N$ (an even integer) qubits. A
Bell inequality involving only two-body correlators with nearest-neighbor
couplings has recently been obtained through a dynamic programming
procedure \cite{Tura2017Energy}:
\begin{eqnarray}
\mathcal{I}_{1} & = & \sum_{k=0}^{N/2-1}(1+\delta)\mathcal{I}_{\text{even}}^{(k)}+(1-\delta)\mathcal{I}_{\text{odd}}^{(k)}\geq\mathcal{B}_{1}^{(c)},\label{BIsr2bc}
\end{eqnarray}
where $\mathcal{I}_{\text{even}}^{(k)}=\sum_{a=0}^{4}\sum_{b=0}^{3}\Lambda_{a,b}(\Delta)\langle\mathcal{M}_{a}^{(2k)}\mathcal{M}_{b}^{(2k+1)}\rangle$
and $\mathcal{I}_{\text{odd}}^{(k)}=\mathcal{I}_{\text{even}}^{(k)}(2k\rightarrow2k+1)$
with $\Lambda(\Delta)$ a four-by-three matrix \cite{MLBnonLa}; $\mathcal{B}_{1}^{(c)}$
is the classical bound depending on the real parameters $\delta$
and $\Delta$: $\mathcal{B}_{1}^{(c)}=-(4+2|\Delta|)N$ for $|\Delta|\leq2$
and $|\delta|\leq1$, and $\mathcal{B}_{1}^{(c)}=-4|\Delta|N$ for
$2\leq|\Delta|\leq3$ and $|\delta|\leq1$. By choosing the measurement
settings properly \cite{supplementmlb}, the Bell operator corresponding
to the above inequality reduces to the following XXZ-type Hamiltonian:
\begin{eqnarray*}
H & = & \sum_{k=0}^{N-1}g_{k}(\delta)[\hat{\sigma}_{k}^{x}\hat{\sigma}_{k+1}^{x}+\hat{\sigma}_{k}^{y}\hat{\sigma}_{k+1}^{y}+\Delta\hat{\sigma}_{k}^{z}\hat{\sigma}_{k+1}^{z}],
\end{eqnarray*}
where $g_{k}(\delta)=4[1+(-1)^{k}]/\sqrt{3}$, and $\hat{\sigma}^{x}$,
$\hat{\sigma}^{y}$ and $\hat{\sigma}^{z}$ are the usual Pauli matrices.
For this particular measurement settings, the maximal quantum violation
of Ineq. (\ref{BIsr2bc}) corresponds to the ground state energy of
$H$ and can be calculated using DMRG, as already discussed in Ref.
\cite{Tura2017Energy}. Here, we use the above introduced reinforcement
learning method to obtain the same quantum violation. 

Our results are plotted in Fig. \ref{fig:Reinforcement-learning-XXZ}.
In Fig. \ref{fig:Reinforcement-learning-XXZ}(a), we compare our results
with that from ED for $N=20$. As shown in this figure, the RBM result
matches the ED result very well \cite{BvioSup}. We find that the
quantum expectation value of $\mathcal{I}_{1}$ (denoted by $\mathcal{Q}_{v}^{(1)}$)
decreases approximately linearly as we increase $\Delta$. There is
a critical value $\Delta\approx2.4$, after which no quantum violation
will be observed. In Fig. \ref{fig:Reinforcement-learning-XXZ}(b),
we show the convergence of the RBM learning and compare the obtained
results with that of DMRG. We find that the initial random RBM states
typically do not violate the Ineq. \ref{BIsr2bc}, but as the learning
process goes on, $\mathcal{Q}_{v}^{(1)}$ will decrease and begin
to violate the inequality after a certain critical iteration number.
As the iteration number increases further, $\mathcal{Q}_{v}^{(1)}$
quickly converges to the DMRG value, validating the effectiveness
of the RBM method. Fig. \ref{fig:Reinforcement-learning-XXZ}(c) shows
the converged $\mathcal{Q}_{v}^{(1)}$ as a function of $N$. We find
that $\mathcal{Q}_{v}^{(1)}$ decreases linearly with increasing $N$
for the chosen parameters $(\delta,\Delta)=(0.9,2)$. For $\Delta=2$,
the slope for $\mathcal{Q}_{v}^{(1)}$ is smaller than that of $\mathcal{B}_{1}^{(c)}$,
thus the larger N the stronger quantum violations. For $\Delta=3$,
no violation is observed for all $N$, which is consistent with the
results in Ref. \cite{Tura2017Energy}.

\begin{figure}
\includegraphics[width=0.491\textwidth]{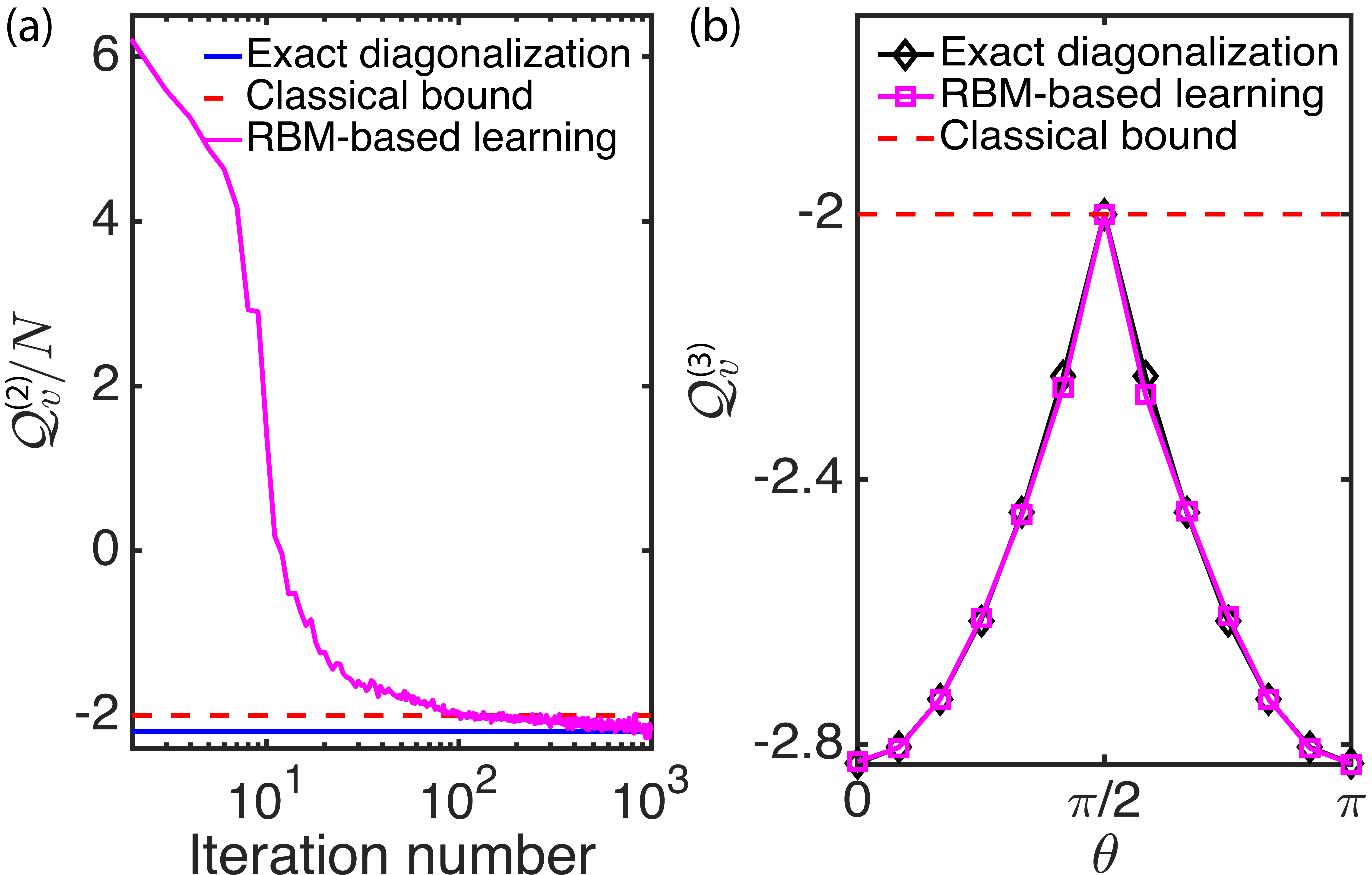}

\caption{(a) RBM-learned quantum expectation value ($\mathcal{Q}_{v}^{(2)}$)
as a function of the iteration number, for a typical random sample
of measurement angles (see \cite{BvioSup}). (b) RBM-learned quantum
violations ($\mathcal{Q}_{v}^{(3)}$) of Ineq. (\ref{eq:BInC}) as
a function of measurement angle $\theta$. In both (a) and (b), the
system size is fixed to be $N=20$. \label{fig:BI2CnC}}
\end{figure}

\textit{Bell inequalities with all-to-all two-body correlators.}\textemdash As
the second example to show the power of RBM in detecting nonlocality,
we consider the following Bell inequality introduced by Tura \textit{et
al} \cite{Tura2014Detecting}, which involves \textit{all-to-all}
two-body correlators and thus are beyond the scope of DMRG:
\begin{eqnarray}
\mathcal{I}_{2} & = & -2S_{0}-S_{01}+\frac{1}{2}(S_{00}+S_{11})\geq\mathcal{B}_{2}^{(c)},\label{eq:BI-All-to-All}
\end{eqnarray}
where the one- and two- body correlators are defined as: $S_{a}=\sum_{k=1}^{N}\langle\mathcal{M}_{a}^{(k)}\rangle$
and $S_{ab}=\sum_{k\neq l}^{N}\langle\mathcal{M}_{a}^{(k)}\mathcal{M}_{b}^{(l)}\rangle$
($a,b=0,1$), and the classical bound $\mathcal{B}_{2}^{(c)}=-2N$.
This inequality has been used in a recent experiment to demonstrate
many-body nonlocality of about $480$ atoms in a Bose-Einstein condensate
\cite{Schmied2016Bell}. For permutationally-symmetric states, its
quantum violations were numerically studied in Ref. \cite{Tura2014Detecting}.
Here, we find that, using the RBM approach, one can obtain the same
maximal violations readily if one chooses a permutation-invariant
neural network. More interestingly, we find that the RBM approach
also works for the cases where the permutation symmetry is released.
To this end, we consider a scenario where the measurement settings
are chosen as: $\mathcal{M}_{0}^{(k)}=\sigma^{z}$ and $\mathcal{M}_{1}^{(k)}=\cos\theta_{k}\sigma^{z}+\sin\theta_{k}\sigma^{x}$
($k=1,2,\cdots,N$), where $\theta_{k}$ are random rotation angles
drawn from some uniform distributions \cite{BvioSup}. We mention
that in a real experiment, the measurement angles will never be exact
due to various control imperfections or system noises. For instance,
in quantum dot spin-qubit experiments, the precision of single qubit
rotations is typically limited due to charge fluctuations \cite{Dial2013Charge}
and Overhauser noise \cite{Medford2012Scaling,Fink2013Noise}. Thus
our consideration of random measurement settings is of both theoretical
and experimental relevance. In Fig. \ref{fig:BI2CnC}(a), we show
the quantum expectation value corresponding to $\mathcal{I}_{2}$
(denoted as $\mathcal{Q}_{v}^{(2)}$) as a function of the iteration
number of the learning process for a typical random sample of $\theta_{k}$s.
It is clear that $\mathcal{Q}_{v}^{(2)}$ decreases as the learning
process continues and becomes smaller than the classical bound at
a critical iteration number. It converges to the exact minimal value
as we increase the iteration number further. 

\textit{Bell inequalities with multipartite correlators.}\textemdash To
show that RBM is also capable of dealing with Bell inequalities with
multipartite correlators, we consider the following Bell inequality
introduced in Ref. \cite{Baccari2017Efficient}:
\begin{eqnarray}
\mathcal{I}_{3} & = & -\langle\mathcal{M}_{0}^{(1)}\mathcal{M}_{0}^{(2)}\cdots\mathcal{M}_{0}^{(N)}\rangle-\langle\mathcal{M}_{1}^{(1)}\mathcal{M}_{0}^{(2)}\cdots\mathcal{M}_{0}^{(N)}\rangle\nonumber \\
 & + & \frac{1}{N-1}\sum_{k=2}^{N}[\langle\mathcal{M}_{0}^{(1)}\mathcal{M}_{1}^{(k)}\rangle-\langle\mathcal{M}_{1}^{(1)}\mathcal{M}_{1}^{(k)}\rangle]\geq-2.\label{eq:BInC}
\end{eqnarray}
This inequality contains only two dichotomic measurements per party,
hence in order to find its maximal quantum violation it is sufficient
to consider only traceless real observables \cite{Masanes2005Extremal,Toner2006Monogamy}.
As a result, we consider the following choice of measurements: $\mathcal{M}_{0}^{(1)}=\hat{\sigma}^{z},$
$\mathcal{M}_{1}^{(1)}=\cos\theta\hat{\sigma}^{x}+\sin\theta\hat{\sigma}^{z}$,
$\mathcal{M}_{0}^{(k)}=\hat{\sigma}^{z}$ and $\mathcal{M}_{1}^{(k)}=\hat{\sigma}^{x}$
for all $k=2,\cdots N$. By using RBM-based reinforcement learning,
we have computed the quantum violations of Ineq. (\ref{eq:BInC})
and part of our results are plotted in Fig. \ref{fig:BI2CnC}(b) \cite{BvioSup}.
From this figure, we find that the Ineq. (\ref{eq:BInC}) is always
violated when $\theta\neq\pi/2$ and the maximal violation is achieved
at $\theta=0$ or $\pi$. When $\theta=\pi/2$, $\mathcal{M}_{0}^{(1)}=\mathcal{M}_{1}^{(1)}=\hat{\sigma}^{z}$
and the first party actually has only one measurement hence no quantum
violation can be obtained. In addition, from our numerical results
we also find that the maximal violation of Ineq. (\ref{eq:BInC})
is always $-2\sqrt{2}$, independent of the system size \cite{BvioSup}.
This can be understood from the observation that the Ineq. (\ref{eq:BInC})
is in fact reminiscent of the Clauser-Horne-Shimony-Holt inequality
\cite{Clauser1969Proposed}, whose maximal quantum violation has proved
to be bounded by $2\sqrt{2}$ \cite{Cirel1980quantum}.

We emphasize that in the last two examples, we did not specify the
spatial dimensionality of the systems. Unlike DMRG, our RBM approach
works for any dimension. In addition, as shown in Ref. \cite{Deng2017Quantum},
entanglement is not a limiting factor for the efficiency of the neural-network
representation of quantum many-body states. Thus, we expect that RBM
can be used to detect many-body nonlocality for quantum states with
massive (e.g., volume-law) entanglement as well. This implies another
unparalleled advantage of the RBM approach, when compared with traditional
methods, such as DMRG, PEPS \cite{Verstraete2008Matrix} (projected
entangled pair states), or MERA \cite{Vidal2008Class} (multiscale
entanglement renormalization ansatz). We also note that one may use
other type of neural networks (e.g., deep Boltzmann machine \cite{Gao2017Efficient}
or feedforward neural networks \cite{Schmidhuber2015Deep}, etc.)
with different learning algorithms to detect many-body nonlocality.
A complete study on detecting nonlocality with different neural network
would not only bring new powerful tools for solving intricate problems
in the quantum information area, but also provide helpful insight
on understanding the internal data structures of the networks themselves.
We leave this interesting and important topic for future investigation. 

\textit{Discussion and conclusion.\textemdash }Finding out experimentally-friendly
Bell inequalities for a given many-body system is a challenging problem,
since in general the complexity of characterizing the set of classical
correlations scales exponentially with the system size. In the future,
it would also be interesting to study how machine learning can provide
valuable ideas in designing optimal Bell inequalities for many-body
systems. Particularly, recent experiments in cold atomic \cite{Bernien2017Probing}
and trapped ion \cite{Zhang2017ObservationDPT} systems have realized
programmable quantum simulators with more than fifty qubits and observed
exotic quantum dynamics and phases transitions. It is highly desirable
to find appropriate Bell inequalities that can be used in these experiments
to demonstrate many-body nonlocality. We believe that machine learning
will provide valuable wisdom in tackling this problem as well. 

In summary, we have introduced machine learning to the detection of
quantum nonlocality in many-body systems. Our discussion is mainly
based on the RBM architecture, but its generalizations to other artificial
neural networks are possible and straightforward. Through three concrete
examples, we have demonstrated that RBM-based reinforcement learning
shows remarkable power in computing quantum violations of certain
multipartite Bell inequalities. Our results open a door for machine
learning Bell nonlocality, which would benefit future studies across
quantum information, machine learning, and artificial intelligence. 

D. L. D acknowledges S. Das Sarma, Lu-Ming Duan, Xiaopeng Li, Jing-Ling
Chen and C. H. Oh for previous collaborations on related works. D.
L. D. thanks Shengtao Wang for helpful discussions. This work is supported
by Laboratory for Physical Sciences and Microsoft. We acknowledge
the University of Maryland supercomputing resources (http://www.it.umd.edu/hpcc)
made available in conducting the research reported in this paper.

\bibliographystyle{/Users/dldeng/Documents/DLDENG/Dropbox/Deng-DataBasis/tex/apsrev4-1-title}
\bibliography{/Users/dldeng/Documents/DLDENG/Dropbox/Deng-DataBasis/tex/Dengbib}

\begin{thebibliography}{99}%
\makeatletter
\providecommand \@ifxundefined [1]{%
 \@ifx{#1\undefined}
}%
\providecommand \@ifnum [1]{%
 \ifnum #1\expandafter \@firstoftwo
 \else \expandafter \@secondoftwo
 \fi
}%
\providecommand \@ifx [1]{%
 \ifx #1\expandafter \@firstoftwo
 \else \expandafter \@secondoftwo
 \fi
}%
\providecommand \natexlab [1]{#1}%
\providecommand \enquote  [1]{``#1''}%
\providecommand \bibnamefont  [1]{#1}%
\providecommand \bibfnamefont [1]{#1}%
\providecommand \citenamefont [1]{#1}%
\providecommand \href@noop [0]{\@secondoftwo}%
\providecommand \href [0]{\begingroup \@sanitize@url \@href}%
\providecommand \@href[1]{\@@startlink{#1}\@@href}%
\providecommand \@@href[1]{\endgroup#1\@@endlink}%
\providecommand \@sanitize@url [0]{\catcode `\\12\catcode `\$12\catcode
  `\&12\catcode `\#12\catcode `\^12\catcode `\_12\catcode `\%12\relax}%
\providecommand \@@startlink[1]{}%
\providecommand \@@endlink[0]{}%
\providecommand \url  [0]{\begingroup\@sanitize@url \@url }%
\providecommand \@url [1]{\endgroup\@href {#1}{\urlprefix }}%
\providecommand \urlprefix  [0]{URL }%
\providecommand \Eprint [0]{\href }%
\providecommand \doibase [0]{http://dx.doi.org/}%
\providecommand \selectlanguage [0]{\@gobble}%
\providecommand \bibinfo  [0]{\@secondoftwo}%
\providecommand \bibfield  [0]{\@secondoftwo}%
\providecommand \translation [1]{[#1]}%
\providecommand \BibitemOpen [0]{}%
\providecommand \bibitemStop [0]{}%
\providecommand \bibitemNoStop [0]{.\EOS\space}%
\providecommand \EOS [0]{\spacefactor3000\relax}%
\providecommand \BibitemShut  [1]{\csname bibitem#1\endcsname}%
\let\auto@bib@innerbib\@empty
\bibitem [{\citenamefont {Bell}(1964)}]{Bell1964OnEPR}%
  \BibitemOpen
  \bibfield  {author} {\bibinfo {author} {\bibfnamefont {J.~S.}\ \bibnamefont
  {Bell}},\ }\bibfield  {title} {\enquote {\bibinfo {title} {On the einstein
  podolsky rosen paradox},}\ }\href@noop {} {\bibfield  {journal} {\bibinfo
  {journal} {(Long Island City, N.Y.)}\ }\textbf {\bibinfo {volume} {1}},\
  \bibinfo {pages} {195} (\bibinfo {year} {1964})}\BibitemShut {NoStop}%
\bibitem [{\citenamefont {Brunner}\ \emph {et~al.}(2014)\citenamefont
  {Brunner}, \citenamefont {Cavalcanti}, \citenamefont {Pironio}, \citenamefont
  {Scarani},\ and\ \citenamefont {Wehner}}]{Brunner2014Bell}%
  \BibitemOpen
  \bibfield  {author} {\bibinfo {author} {\bibfnamefont {N.}~\bibnamefont
  {Brunner}}, \bibinfo {author} {\bibfnamefont {D.}~\bibnamefont {Cavalcanti}},
  \bibinfo {author} {\bibfnamefont {S.}~\bibnamefont {Pironio}}, \bibinfo
  {author} {\bibfnamefont {V.}~\bibnamefont {Scarani}}, \ and\ \bibinfo
  {author} {\bibfnamefont {S.}~\bibnamefont {Wehner}},\ }\bibfield  {title}
  {\enquote {\bibinfo {title} {Bell nonlocality},}\ }\href {\doibase
  10.1103/RevModPhys.86.419} {\bibfield  {journal} {\bibinfo  {journal} {Rev.
  Mod. Phys.}\ }\textbf {\bibinfo {volume} {86}},\ \bibinfo {pages} {419}
  (\bibinfo {year} {2014})}\BibitemShut {NoStop}%
\bibitem [{\citenamefont {Freedman}\ and\ \citenamefont
  {Clauser}(1972)}]{Freedman1972Experimental}%
  \BibitemOpen
  \bibfield  {author} {\bibinfo {author} {\bibfnamefont {S.~J.}\ \bibnamefont
  {Freedman}}\ and\ \bibinfo {author} {\bibfnamefont {J.~F.}\ \bibnamefont
  {Clauser}},\ }\bibfield  {title} {\enquote {\bibinfo {title} {Experimental
  test of local hidden-variable theories},}\ }\href {\doibase
  10.1103/PhysRevLett.28.938} {\bibfield  {journal} {\bibinfo  {journal} {Phys.
  Rev. Lett.}\ }\textbf {\bibinfo {volume} {28}},\ \bibinfo {pages} {938}
  (\bibinfo {year} {1972})}\BibitemShut {NoStop}%
\bibitem [{\citenamefont {Aspect}\ \emph {et~al.}(1982)\citenamefont {Aspect},
  \citenamefont {Dalibard},\ and\ \citenamefont
  {Roger}}]{Aspect1982Experimental}%
  \BibitemOpen
  \bibfield  {author} {\bibinfo {author} {\bibfnamefont {A.}~\bibnamefont
  {Aspect}}, \bibinfo {author} {\bibfnamefont {J.}~\bibnamefont {Dalibard}}, \
  and\ \bibinfo {author} {\bibfnamefont {G.}~\bibnamefont {Roger}},\ }\bibfield
   {title} {\enquote {\bibinfo {title} {Experimental test of bell's
  inequalities using time-varying analyzers},}\ }\href {\doibase
  10.1103/PhysRevLett.49.1804} {\bibfield  {journal} {\bibinfo  {journal}
  {Phys. Rev. Lett.}\ }\textbf {\bibinfo {volume} {49}},\ \bibinfo {pages}
  {1804} (\bibinfo {year} {1982})}\BibitemShut {NoStop}%
\bibitem [{\citenamefont {Weihs}\ \emph {et~al.}(1998)\citenamefont {Weihs},
  \citenamefont {Jennewein}, \citenamefont {Simon}, \citenamefont
  {Weinfurter},\ and\ \citenamefont {Zeilinger}}]{Weihs1998Violation}%
  \BibitemOpen
  \bibfield  {author} {\bibinfo {author} {\bibfnamefont {G.}~\bibnamefont
  {Weihs}}, \bibinfo {author} {\bibfnamefont {T.}~\bibnamefont {Jennewein}},
  \bibinfo {author} {\bibfnamefont {C.}~\bibnamefont {Simon}}, \bibinfo
  {author} {\bibfnamefont {H.}~\bibnamefont {Weinfurter}}, \ and\ \bibinfo
  {author} {\bibfnamefont {A.}~\bibnamefont {Zeilinger}},\ }\bibfield  {title}
  {\enquote {\bibinfo {title} {Violation of bell's inequality under strict
  einstein locality conditions},}\ }\href {\doibase
  10.1103/PhysRevLett.81.5039} {\bibfield  {journal} {\bibinfo  {journal}
  {Phys. Rev. Lett.}\ }\textbf {\bibinfo {volume} {81}},\ \bibinfo {pages}
  {5039} (\bibinfo {year} {1998})}\BibitemShut {NoStop}%
\bibitem [{\citenamefont {Rowe}\ \emph {et~al.}(2001)\citenamefont {Rowe},
  \citenamefont {Kielpinski}, \citenamefont {Meyer}, \citenamefont {Sackett}
  \emph {et~al.}}]{Rowe2001Experimental}%
  \BibitemOpen
  \bibfield  {author} {\bibinfo {author} {\bibfnamefont {M.~A.}\ \bibnamefont
  {Rowe}}, \bibinfo {author} {\bibfnamefont {D.}~\bibnamefont {Kielpinski}},
  \bibinfo {author} {\bibfnamefont {V.}~\bibnamefont {Meyer}}, \bibinfo
  {author} {\bibfnamefont {C.~A.}\ \bibnamefont {Sackett}},  \emph {et~al.},\
  }\bibfield  {title} {\enquote {\bibinfo {title} {Experimental violation of a
  bell's inequality with efficient detection},}\ }\href {\doibase
  10.1038/35057215} {\bibfield  {journal} {\bibinfo  {journal} {Nature}\
  }\textbf {\bibinfo {volume} {409}},\ \bibinfo {pages} {791} (\bibinfo {year}
  {2001})}\BibitemShut {NoStop}%
\bibitem [{\citenamefont {Giustina}\ \emph {et~al.}(2013)\citenamefont
  {Giustina}, \citenamefont {Mech}, \citenamefont {Ramelow}, \citenamefont
  {Wittmann}, \citenamefont {Kofler}, \citenamefont {Beyer}, \citenamefont
  {Lita}, \citenamefont {Calkins}, \citenamefont {Gerrits}, \citenamefont {Nam}
  \emph {et~al.}}]{Giustina2013Bell}%
  \BibitemOpen
  \bibfield  {author} {\bibinfo {author} {\bibfnamefont {M.}~\bibnamefont
  {Giustina}}, \bibinfo {author} {\bibfnamefont {A.}~\bibnamefont {Mech}},
  \bibinfo {author} {\bibfnamefont {S.}~\bibnamefont {Ramelow}}, \bibinfo
  {author} {\bibfnamefont {B.}~\bibnamefont {Wittmann}}, \bibinfo {author}
  {\bibfnamefont {J.}~\bibnamefont {Kofler}}, \bibinfo {author} {\bibfnamefont
  {J.}~\bibnamefont {Beyer}}, \bibinfo {author} {\bibfnamefont
  {A.}~\bibnamefont {Lita}}, \bibinfo {author} {\bibfnamefont {B.}~\bibnamefont
  {Calkins}}, \bibinfo {author} {\bibfnamefont {T.}~\bibnamefont {Gerrits}},
  \bibinfo {author} {\bibfnamefont {S.~W.}\ \bibnamefont {Nam}},  \emph
  {et~al.},\ }\bibfield  {title} {\enquote {\bibinfo {title} {Bell violation
  using entangled photons without the fair-sampling assumption},}\ }\href
  {\doibase 10.1038/nature12012} {\bibfield  {journal} {\bibinfo  {journal}
  {Nature}\ }\textbf {\bibinfo {volume} {497}},\ \bibinfo {pages} {227}
  (\bibinfo {year} {2013})}\BibitemShut {NoStop}%
\bibitem [{\citenamefont {Christensen}\ \emph {et~al.}(2013)\citenamefont
  {Christensen}, \citenamefont {McCusker}, \citenamefont {Altepeter},
  \citenamefont {Calkins}, \citenamefont {Gerrits}, \citenamefont {Lita},
  \citenamefont {Miller}, \citenamefont {Shalm}, \citenamefont {Zhang},
  \citenamefont {Nam}, \citenamefont {Brunner}, \citenamefont {Lim},
  \citenamefont {Gisin},\ and\ \citenamefont
  {Kwiat}}]{Christensen2013Detection}%
  \BibitemOpen
  \bibfield  {author} {\bibinfo {author} {\bibfnamefont {B.~G.}\ \bibnamefont
  {Christensen}}, \bibinfo {author} {\bibfnamefont {K.~T.}\ \bibnamefont
  {McCusker}}, \bibinfo {author} {\bibfnamefont {J.~B.}\ \bibnamefont
  {Altepeter}}, \bibinfo {author} {\bibfnamefont {B.}~\bibnamefont {Calkins}},
  \bibinfo {author} {\bibfnamefont {T.}~\bibnamefont {Gerrits}}, \bibinfo
  {author} {\bibfnamefont {A.~E.}\ \bibnamefont {Lita}}, \bibinfo {author}
  {\bibfnamefont {A.}~\bibnamefont {Miller}}, \bibinfo {author} {\bibfnamefont
  {L.~K.}\ \bibnamefont {Shalm}}, \bibinfo {author} {\bibfnamefont
  {Y.}~\bibnamefont {Zhang}}, \bibinfo {author} {\bibfnamefont {S.~W.}\
  \bibnamefont {Nam}}, \bibinfo {author} {\bibfnamefont {N.}~\bibnamefont
  {Brunner}}, \bibinfo {author} {\bibfnamefont {C.~C.~W.}\ \bibnamefont {Lim}},
  \bibinfo {author} {\bibfnamefont {N.}~\bibnamefont {Gisin}}, \ and\ \bibinfo
  {author} {\bibfnamefont {P.~G.}\ \bibnamefont {Kwiat}},\ }\bibfield  {title}
  {\enquote {\bibinfo {title} {Detection-loophole-free test of quantum
  nonlocality, and applications},}\ }\href {\doibase
  10.1103/PhysRevLett.111.130406} {\bibfield  {journal} {\bibinfo  {journal}
  {Phys. Rev. Lett.}\ }\textbf {\bibinfo {volume} {111}},\ \bibinfo {pages}
  {130406} (\bibinfo {year} {2013})}\BibitemShut {NoStop}%
\bibitem [{\citenamefont {Eibl}\ \emph {et~al.}(2003)\citenamefont {Eibl},
  \citenamefont {Gaertner}, \citenamefont {Bourennane}, \citenamefont
  {Kurtsiefer}, \citenamefont {\ifmmode~\dot{Z}\else \.{Z}\fi{}ukowski},\ and\
  \citenamefont {Weinfurter}}]{Eibl2003Experimental}%
  \BibitemOpen
  \bibfield  {author} {\bibinfo {author} {\bibfnamefont {M.}~\bibnamefont
  {Eibl}}, \bibinfo {author} {\bibfnamefont {S.}~\bibnamefont {Gaertner}},
  \bibinfo {author} {\bibfnamefont {M.}~\bibnamefont {Bourennane}}, \bibinfo
  {author} {\bibfnamefont {C.}~\bibnamefont {Kurtsiefer}}, \bibinfo {author}
  {\bibfnamefont {M.}~\bibnamefont {\ifmmode~\dot{Z}\else \.{Z}\fi{}ukowski}},
  \ and\ \bibinfo {author} {\bibfnamefont {H.}~\bibnamefont {Weinfurter}},\
  }\bibfield  {title} {\enquote {\bibinfo {title} {Experimental observation of
  four-photon entanglement from parametric down-conversion},}\ }\href {\doibase
  10.1103/PhysRevLett.90.200403} {\bibfield  {journal} {\bibinfo  {journal}
  {Phys. Rev. Lett.}\ }\textbf {\bibinfo {volume} {90}},\ \bibinfo {pages}
  {200403} (\bibinfo {year} {2003})}\BibitemShut {NoStop}%
\bibitem [{\citenamefont {Zhao}\ \emph {et~al.}(2003)\citenamefont {Zhao},
  \citenamefont {Yang}, \citenamefont {Chen}, \citenamefont {Zhang},
  \citenamefont {\ifmmode~\dot{Z}\else \.{Z}\fi{}ukowski},\ and\ \citenamefont
  {Pan}}]{Zhao2003Experimental}%
  \BibitemOpen
  \bibfield  {author} {\bibinfo {author} {\bibfnamefont {Z.}~\bibnamefont
  {Zhao}}, \bibinfo {author} {\bibfnamefont {T.}~\bibnamefont {Yang}}, \bibinfo
  {author} {\bibfnamefont {Y.-A.}\ \bibnamefont {Chen}}, \bibinfo {author}
  {\bibfnamefont {A.-N.}\ \bibnamefont {Zhang}}, \bibinfo {author}
  {\bibfnamefont {M.}~\bibnamefont {\ifmmode~\dot{Z}\else \.{Z}\fi{}ukowski}},
  \ and\ \bibinfo {author} {\bibfnamefont {J.-W.}\ \bibnamefont {Pan}},\
  }\bibfield  {title} {\enquote {\bibinfo {title} {Experimental violation of
  local realism by four-photon greenberger-horne-zeilinger entanglement},}\
  }\href {\doibase 10.1103/PhysRevLett.91.180401} {\bibfield  {journal}
  {\bibinfo  {journal} {Phys. Rev. Lett.}\ }\textbf {\bibinfo {volume} {91}},\
  \bibinfo {pages} {180401} (\bibinfo {year} {2003})}\BibitemShut {NoStop}%
\bibitem [{\citenamefont {Lanyon}\ \emph {et~al.}(2014)\citenamefont {Lanyon},
  \citenamefont {Zwerger}, \citenamefont {Jurcevic}, \citenamefont {Hempel},
  \citenamefont {D\"ur}, \citenamefont {Briegel}, \citenamefont {Blatt},\ and\
  \citenamefont {Roos}}]{Lanyon2014Experimental}%
  \BibitemOpen
  \bibfield  {author} {\bibinfo {author} {\bibfnamefont {B.~P.}\ \bibnamefont
  {Lanyon}}, \bibinfo {author} {\bibfnamefont {M.}~\bibnamefont {Zwerger}},
  \bibinfo {author} {\bibfnamefont {P.}~\bibnamefont {Jurcevic}}, \bibinfo
  {author} {\bibfnamefont {C.}~\bibnamefont {Hempel}}, \bibinfo {author}
  {\bibfnamefont {W.}~\bibnamefont {D\"ur}}, \bibinfo {author} {\bibfnamefont
  {H.~J.}\ \bibnamefont {Briegel}}, \bibinfo {author} {\bibfnamefont
  {R.}~\bibnamefont {Blatt}}, \ and\ \bibinfo {author} {\bibfnamefont {C.~F.}\
  \bibnamefont {Roos}},\ }\bibfield  {title} {\enquote {\bibinfo {title}
  {Experimental violation of multipartite bell inequalities with trapped
  ions},}\ }\href {\doibase 10.1103/PhysRevLett.112.100403} {\bibfield
  {journal} {\bibinfo  {journal} {Phys. Rev. Lett.}\ }\textbf {\bibinfo
  {volume} {112}},\ \bibinfo {pages} {100403} (\bibinfo {year}
  {2014})}\BibitemShut {NoStop}%
\bibitem [{\citenamefont {Hofmann}\ \emph {et~al.}(2012)\citenamefont
  {Hofmann}, \citenamefont {Krug}, \citenamefont {Ortegel}, \citenamefont
  {G{\'e}rard}, \citenamefont {Weber}, \citenamefont {Rosenfeld},\ and\
  \citenamefont {Weinfurter}}]{Hofmann2012Heralded}%
  \BibitemOpen
  \bibfield  {author} {\bibinfo {author} {\bibfnamefont {J.}~\bibnamefont
  {Hofmann}}, \bibinfo {author} {\bibfnamefont {M.}~\bibnamefont {Krug}},
  \bibinfo {author} {\bibfnamefont {N.}~\bibnamefont {Ortegel}}, \bibinfo
  {author} {\bibfnamefont {L.}~\bibnamefont {G{\'e}rard}}, \bibinfo {author}
  {\bibfnamefont {M.}~\bibnamefont {Weber}}, \bibinfo {author} {\bibfnamefont
  {W.}~\bibnamefont {Rosenfeld}}, \ and\ \bibinfo {author} {\bibfnamefont
  {H.}~\bibnamefont {Weinfurter}},\ }\bibfield  {title} {\enquote {\bibinfo
  {title} {Heralded entanglement between widely separated atoms},}\ }\href@noop
  {} {\bibfield  {journal} {\bibinfo  {journal} {Science}\ }\textbf {\bibinfo
  {volume} {337}},\ \bibinfo {pages} {72} (\bibinfo {year} {2012})}\BibitemShut
  {NoStop}%
\bibitem [{\citenamefont {Pfaff}\ \emph {et~al.}(2013)\citenamefont {Pfaff},
  \citenamefont {Taminiau}, \citenamefont {Robledo}, \citenamefont {Bernien},
  \citenamefont {Markham}, \citenamefont {Twitchen},\ and\ \citenamefont
  {Hanson}}]{Pfaff2013Demonstration}%
  \BibitemOpen
  \bibfield  {author} {\bibinfo {author} {\bibfnamefont {W.}~\bibnamefont
  {Pfaff}}, \bibinfo {author} {\bibfnamefont {T.~H.}\ \bibnamefont {Taminiau}},
  \bibinfo {author} {\bibfnamefont {L.}~\bibnamefont {Robledo}}, \bibinfo
  {author} {\bibfnamefont {H.}~\bibnamefont {Bernien}}, \bibinfo {author}
  {\bibfnamefont {M.}~\bibnamefont {Markham}}, \bibinfo {author} {\bibfnamefont
  {D.~J.}\ \bibnamefont {Twitchen}}, \ and\ \bibinfo {author} {\bibfnamefont
  {R.}~\bibnamefont {Hanson}},\ }\bibfield  {title} {\enquote {\bibinfo {title}
  {Demonstration of entanglement-by-measurement of solid-state qubits},}\
  }\href {\doibase 10.1038/nphys2444} {\bibfield  {journal} {\bibinfo
  {journal} {Nat. Phys.}\ }\textbf {\bibinfo {volume} {9}} (\bibinfo {year}
  {2013}),\ 10.1038/nphys2444}\BibitemShut {NoStop}%
\bibitem [{\citenamefont {Ansmann}\ \emph {et~al.}(2009)\citenamefont
  {Ansmann}, \citenamefont {Wang}, \citenamefont {Bialczak}, \citenamefont
  {Hofheinz}, \citenamefont {Lucero}, \citenamefont {Neeley}, \citenamefont
  {O'connell}, \citenamefont {Sank}, \citenamefont {Weides}, \citenamefont
  {Wenner} \emph {et~al.}}]{Ansmann2009Violation}%
  \BibitemOpen
  \bibfield  {author} {\bibinfo {author} {\bibfnamefont {M.}~\bibnamefont
  {Ansmann}}, \bibinfo {author} {\bibfnamefont {H.}~\bibnamefont {Wang}},
  \bibinfo {author} {\bibfnamefont {R.~C.}\ \bibnamefont {Bialczak}}, \bibinfo
  {author} {\bibfnamefont {M.}~\bibnamefont {Hofheinz}}, \bibinfo {author}
  {\bibfnamefont {E.}~\bibnamefont {Lucero}}, \bibinfo {author} {\bibfnamefont
  {M.}~\bibnamefont {Neeley}}, \bibinfo {author} {\bibfnamefont
  {A.}~\bibnamefont {O'connell}}, \bibinfo {author} {\bibfnamefont
  {D.}~\bibnamefont {Sank}}, \bibinfo {author} {\bibfnamefont {M.}~\bibnamefont
  {Weides}}, \bibinfo {author} {\bibfnamefont {J.}~\bibnamefont {Wenner}},
  \emph {et~al.},\ }\bibfield  {title} {\enquote {\bibinfo {title} {Violation
  of bell's inequality in josephson phase qubits},}\ }\href {\doibase
  10.1038/nature08363} {\bibfield  {journal} {\bibinfo  {journal} {Nature}\
  }\textbf {\bibinfo {volume} {461}},\ \bibinfo {pages} {504} (\bibinfo {year}
  {2009})}\BibitemShut {NoStop}%
\bibitem [{\citenamefont {Hensen}\ \emph {et~al.}(2015)\citenamefont {Hensen},
  \citenamefont {Bernien}, \citenamefont {Dr{\'e}au}, \citenamefont {Reiserer},
  \citenamefont {Kalb}, \citenamefont {Blok}, \citenamefont {Ruitenberg},
  \citenamefont {Vermeulen}, \citenamefont {Schouten}, \citenamefont
  {Abell{\'a}n} \emph {et~al.}}]{Hensen2015Loophole}%
  \BibitemOpen
  \bibfield  {author} {\bibinfo {author} {\bibfnamefont {B.}~\bibnamefont
  {Hensen}}, \bibinfo {author} {\bibfnamefont {H.}~\bibnamefont {Bernien}},
  \bibinfo {author} {\bibfnamefont {A.~E.}\ \bibnamefont {Dr{\'e}au}}, \bibinfo
  {author} {\bibfnamefont {A.}~\bibnamefont {Reiserer}}, \bibinfo {author}
  {\bibfnamefont {N.}~\bibnamefont {Kalb}}, \bibinfo {author} {\bibfnamefont
  {M.~S.}\ \bibnamefont {Blok}}, \bibinfo {author} {\bibfnamefont
  {J.}~\bibnamefont {Ruitenberg}}, \bibinfo {author} {\bibfnamefont {R.~F.}\
  \bibnamefont {Vermeulen}}, \bibinfo {author} {\bibfnamefont {R.~N.}\
  \bibnamefont {Schouten}}, \bibinfo {author} {\bibfnamefont {C.}~\bibnamefont
  {Abell{\'a}n}},  \emph {et~al.},\ }\bibfield  {title} {\enquote {\bibinfo
  {title} {Loophole-free bell inequality violation using electron spins
  separated by 1.3 kilometres},}\ }\href {\doibase 10.1038/nature15759}
  {\bibfield  {journal} {\bibinfo  {journal} {Nature}\ }\textbf {\bibinfo
  {volume} {526}},\ \bibinfo {pages} {682} (\bibinfo {year}
  {2015})}\BibitemShut {NoStop}%
\bibitem [{\citenamefont {Giustina}\ \emph {et~al.}(2015)\citenamefont
  {Giustina}, \citenamefont {Versteegh}, \citenamefont {Wengerowsky},
  \citenamefont {Handsteiner}, \citenamefont {Hochrainer}, \citenamefont
  {Phelan}, \citenamefont {Steinlechner}, \citenamefont {Kofler}, \citenamefont
  {Larsson}, \citenamefont {Abell\'an}, \citenamefont {Amaya}, \citenamefont
  {Pruneri}, \citenamefont {Mitchell}, \citenamefont {Beyer}, \citenamefont
  {Gerrits}, \citenamefont {Lita}, \citenamefont {Shalm}, \citenamefont {Nam},
  \citenamefont {Scheidl}, \citenamefont {Ursin}, \citenamefont {Wittmann},\
  and\ \citenamefont {Zeilinger}}]{Giustina2015Significant}%
  \BibitemOpen
  \bibfield  {author} {\bibinfo {author} {\bibfnamefont {M.}~\bibnamefont
  {Giustina}}, \bibinfo {author} {\bibfnamefont {M.~A.~M.}\ \bibnamefont
  {Versteegh}}, \bibinfo {author} {\bibfnamefont {S.}~\bibnamefont
  {Wengerowsky}}, \bibinfo {author} {\bibfnamefont {J.}~\bibnamefont
  {Handsteiner}}, \bibinfo {author} {\bibfnamefont {A.}~\bibnamefont
  {Hochrainer}}, \bibinfo {author} {\bibfnamefont {K.}~\bibnamefont {Phelan}},
  \bibinfo {author} {\bibfnamefont {F.}~\bibnamefont {Steinlechner}}, \bibinfo
  {author} {\bibfnamefont {J.}~\bibnamefont {Kofler}}, \bibinfo {author}
  {\bibfnamefont {J.-A.}\ \bibnamefont {Larsson}}, \bibinfo {author}
  {\bibfnamefont {C.}~\bibnamefont {Abell\'an}}, \bibinfo {author}
  {\bibfnamefont {W.}~\bibnamefont {Amaya}}, \bibinfo {author} {\bibfnamefont
  {V.}~\bibnamefont {Pruneri}}, \bibinfo {author} {\bibfnamefont {M.~W.}\
  \bibnamefont {Mitchell}}, \bibinfo {author} {\bibfnamefont {J.}~\bibnamefont
  {Beyer}}, \bibinfo {author} {\bibfnamefont {T.}~\bibnamefont {Gerrits}},
  \bibinfo {author} {\bibfnamefont {A.~E.}\ \bibnamefont {Lita}}, \bibinfo
  {author} {\bibfnamefont {L.~K.}\ \bibnamefont {Shalm}}, \bibinfo {author}
  {\bibfnamefont {S.~W.}\ \bibnamefont {Nam}}, \bibinfo {author} {\bibfnamefont
  {T.}~\bibnamefont {Scheidl}}, \bibinfo {author} {\bibfnamefont
  {R.}~\bibnamefont {Ursin}}, \bibinfo {author} {\bibfnamefont
  {B.}~\bibnamefont {Wittmann}}, \ and\ \bibinfo {author} {\bibfnamefont
  {A.}~\bibnamefont {Zeilinger}},\ }\bibfield  {title} {\enquote {\bibinfo
  {title} {Significant-loophole-free test of bell's theorem with entangled
  photons},}\ }\href {\doibase 10.1103/PhysRevLett.115.250401} {\bibfield
  {journal} {\bibinfo  {journal} {Phys. Rev. Lett.}\ }\textbf {\bibinfo
  {volume} {115}},\ \bibinfo {pages} {250401} (\bibinfo {year}
  {2015})}\BibitemShut {NoStop}%
\bibitem [{\citenamefont {Shalm}\ \emph {et~al.}(2015)\citenamefont {Shalm},
  \citenamefont {Meyer-Scott}, \citenamefont {Christensen}, \citenamefont
  {Bierhorst}, \citenamefont {Wayne}, \citenamefont {Stevens}, \citenamefont
  {Gerrits}, \citenamefont {Glancy}, \citenamefont {Hamel}, \citenamefont
  {Allman}, \citenamefont {Coakley}, \citenamefont {Dyer}, \citenamefont
  {Hodge}, \citenamefont {Lita}, \citenamefont {Verma}, \citenamefont
  {Lambrocco}, \citenamefont {Tortorici}, \citenamefont {Migdall},
  \citenamefont {Zhang}, \citenamefont {Kumor}, \citenamefont {Farr},
  \citenamefont {Marsili}, \citenamefont {Shaw}, \citenamefont {Stern},
  \citenamefont {Abell\'an}, \citenamefont {Amaya}, \citenamefont {Pruneri},
  \citenamefont {Jennewein}, \citenamefont {Mitchell}, \citenamefont {Kwiat},
  \citenamefont {Bienfang}, \citenamefont {Mirin}, \citenamefont {Knill},\ and\
  \citenamefont {Nam}}]{Shalm2015Strong}%
  \BibitemOpen
  \bibfield  {author} {\bibinfo {author} {\bibfnamefont {L.~K.}\ \bibnamefont
  {Shalm}}, \bibinfo {author} {\bibfnamefont {E.}~\bibnamefont {Meyer-Scott}},
  \bibinfo {author} {\bibfnamefont {B.~G.}\ \bibnamefont {Christensen}},
  \bibinfo {author} {\bibfnamefont {P.}~\bibnamefont {Bierhorst}}, \bibinfo
  {author} {\bibfnamefont {M.~A.}\ \bibnamefont {Wayne}}, \bibinfo {author}
  {\bibfnamefont {M.~J.}\ \bibnamefont {Stevens}}, \bibinfo {author}
  {\bibfnamefont {T.}~\bibnamefont {Gerrits}}, \bibinfo {author} {\bibfnamefont
  {S.}~\bibnamefont {Glancy}}, \bibinfo {author} {\bibfnamefont {D.~R.}\
  \bibnamefont {Hamel}}, \bibinfo {author} {\bibfnamefont {M.~S.}\ \bibnamefont
  {Allman}}, \bibinfo {author} {\bibfnamefont {K.~J.}\ \bibnamefont {Coakley}},
  \bibinfo {author} {\bibfnamefont {S.~D.}\ \bibnamefont {Dyer}}, \bibinfo
  {author} {\bibfnamefont {C.}~\bibnamefont {Hodge}}, \bibinfo {author}
  {\bibfnamefont {A.~E.}\ \bibnamefont {Lita}}, \bibinfo {author}
  {\bibfnamefont {V.~B.}\ \bibnamefont {Verma}}, \bibinfo {author}
  {\bibfnamefont {C.}~\bibnamefont {Lambrocco}}, \bibinfo {author}
  {\bibfnamefont {E.}~\bibnamefont {Tortorici}}, \bibinfo {author}
  {\bibfnamefont {A.~L.}\ \bibnamefont {Migdall}}, \bibinfo {author}
  {\bibfnamefont {Y.}~\bibnamefont {Zhang}}, \bibinfo {author} {\bibfnamefont
  {D.~R.}\ \bibnamefont {Kumor}}, \bibinfo {author} {\bibfnamefont {W.~H.}\
  \bibnamefont {Farr}}, \bibinfo {author} {\bibfnamefont {F.}~\bibnamefont
  {Marsili}}, \bibinfo {author} {\bibfnamefont {M.~D.}\ \bibnamefont {Shaw}},
  \bibinfo {author} {\bibfnamefont {J.~A.}\ \bibnamefont {Stern}}, \bibinfo
  {author} {\bibfnamefont {C.}~\bibnamefont {Abell\'an}}, \bibinfo {author}
  {\bibfnamefont {W.}~\bibnamefont {Amaya}}, \bibinfo {author} {\bibfnamefont
  {V.}~\bibnamefont {Pruneri}}, \bibinfo {author} {\bibfnamefont
  {T.}~\bibnamefont {Jennewein}}, \bibinfo {author} {\bibfnamefont {M.~W.}\
  \bibnamefont {Mitchell}}, \bibinfo {author} {\bibfnamefont {P.~G.}\
  \bibnamefont {Kwiat}}, \bibinfo {author} {\bibfnamefont {J.~C.}\ \bibnamefont
  {Bienfang}}, \bibinfo {author} {\bibfnamefont {R.~P.}\ \bibnamefont {Mirin}},
  \bibinfo {author} {\bibfnamefont {E.}~\bibnamefont {Knill}}, \ and\ \bibinfo
  {author} {\bibfnamefont {S.~W.}\ \bibnamefont {Nam}},\ }\bibfield  {title}
  {\enquote {\bibinfo {title} {Strong loophole-free test of local realism},}\
  }\href {\doibase 10.1103/PhysRevLett.115.250402} {\bibfield  {journal}
  {\bibinfo  {journal} {Phys. Rev. Lett.}\ }\textbf {\bibinfo {volume} {115}},\
  \bibinfo {pages} {250402} (\bibinfo {year} {2015})}\BibitemShut {NoStop}%
\bibitem [{\citenamefont {Zu}\ \emph {et~al.}(2013)\citenamefont {Zu},
  \citenamefont {Deng}, \citenamefont {Hou}, \citenamefont {Chang},
  \citenamefont {Wang},\ and\ \citenamefont {Duan}}]{Zu2013Experimental}%
  \BibitemOpen
  \bibfield  {author} {\bibinfo {author} {\bibfnamefont {C.}~\bibnamefont
  {Zu}}, \bibinfo {author} {\bibfnamefont {D.-L.}\ \bibnamefont {Deng}},
  \bibinfo {author} {\bibfnamefont {P.-Y.}\ \bibnamefont {Hou}}, \bibinfo
  {author} {\bibfnamefont {X.-Y.}\ \bibnamefont {Chang}}, \bibinfo {author}
  {\bibfnamefont {F.}~\bibnamefont {Wang}}, \ and\ \bibinfo {author}
  {\bibfnamefont {L.-M.}\ \bibnamefont {Duan}},\ }\bibfield  {title} {\enquote
  {\bibinfo {title} {Experimental distillation of quantum nonlocality},}\
  }\href {\doibase 10.1103/PhysRevLett.111.050405} {\bibfield  {journal}
  {\bibinfo  {journal} {Phys. Rev. Lett.}\ }\textbf {\bibinfo {volume} {111}},\
  \bibinfo {pages} {050405} (\bibinfo {year} {2013})}\BibitemShut {NoStop}%
\bibitem [{\citenamefont {Ac\'{\i}n}\ \emph {et~al.}(2007)\citenamefont
  {Ac\'{\i}n}, \citenamefont {Brunner}, \citenamefont {Gisin}, \citenamefont
  {Massar}, \citenamefont {Pironio},\ and\ \citenamefont
  {Scarani}}]{Acin2007Device}%
  \BibitemOpen
  \bibfield  {author} {\bibinfo {author} {\bibfnamefont {A.}~\bibnamefont
  {Ac\'{\i}n}}, \bibinfo {author} {\bibfnamefont {N.}~\bibnamefont {Brunner}},
  \bibinfo {author} {\bibfnamefont {N.}~\bibnamefont {Gisin}}, \bibinfo
  {author} {\bibfnamefont {S.}~\bibnamefont {Massar}}, \bibinfo {author}
  {\bibfnamefont {S.}~\bibnamefont {Pironio}}, \ and\ \bibinfo {author}
  {\bibfnamefont {V.}~\bibnamefont {Scarani}},\ }\bibfield  {title} {\enquote
  {\bibinfo {title} {Device-independent security of quantum cryptography
  against collective attacks},}\ }\href {\doibase
  10.1103/PhysRevLett.98.230501} {\bibfield  {journal} {\bibinfo  {journal}
  {Phys. Rev. Lett.}\ }\textbf {\bibinfo {volume} {98}},\ \bibinfo {pages}
  {230501} (\bibinfo {year} {2007})}\BibitemShut {NoStop}%
\bibitem [{\citenamefont {Pironio}\ and\ \citenamefont
  {Massar}(2013)}]{Pironio2013Security}%
  \BibitemOpen
  \bibfield  {author} {\bibinfo {author} {\bibfnamefont {S.}~\bibnamefont
  {Pironio}}\ and\ \bibinfo {author} {\bibfnamefont {S.}~\bibnamefont
  {Massar}},\ }\bibfield  {title} {\enquote {\bibinfo {title} {Security of
  practical private randomness generation},}\ }\href {\doibase
  10.1103/PhysRevA.87.012336} {\bibfield  {journal} {\bibinfo  {journal} {Phys.
  Rev. A}\ }\textbf {\bibinfo {volume} {87}},\ \bibinfo {pages} {012336}
  (\bibinfo {year} {2013})}\BibitemShut {NoStop}%
\bibitem [{\citenamefont {Vazirani}\ and\ \citenamefont
  {Vidick}(2014)}]{Vazirani2014Fully}%
  \BibitemOpen
  \bibfield  {author} {\bibinfo {author} {\bibfnamefont {U.}~\bibnamefont
  {Vazirani}}\ and\ \bibinfo {author} {\bibfnamefont {T.}~\bibnamefont
  {Vidick}},\ }\bibfield  {title} {\enquote {\bibinfo {title} {Fully
  device-independent quantum key distribution},}\ }\href {\doibase
  10.1103/PhysRevLett.113.140501} {\bibfield  {journal} {\bibinfo  {journal}
  {Phys. Rev. Lett.}\ }\textbf {\bibinfo {volume} {113}},\ \bibinfo {pages}
  {140501} (\bibinfo {year} {2014})}\BibitemShut {NoStop}%
\bibitem [{\citenamefont {Colbeck}(2007)}]{Colbeck2007Quantum}%
  \BibitemOpen
  \bibfield  {author} {\bibinfo {author} {\bibfnamefont {R.}~\bibnamefont
  {Colbeck}},\ }\href@noop {} {\emph {\bibinfo {title} {Quantum And
  Relativistic Protocols For Secure Multi-Party Computation}}}\ (\bibinfo
  {publisher} {PhD thesis, University of Cambridge},\ \bibinfo {year}
  {2007})\BibitemShut {NoStop}%
\bibitem [{\citenamefont {Pironio}\ \emph {et~al.}(2010)\citenamefont
  {Pironio}, \citenamefont {Ac{\'i}n}, \citenamefont {Massar}, \citenamefont
  {de~La~Giroday}, \citenamefont {Matsukevich}, \citenamefont {Maunz},
  \citenamefont {Olmschenk}, \citenamefont {Hayes}, \citenamefont {Luo},
  \citenamefont {Manning} \emph {et~al.}}]{Pironio2010Random}%
  \BibitemOpen
  \bibfield  {author} {\bibinfo {author} {\bibfnamefont {S.}~\bibnamefont
  {Pironio}}, \bibinfo {author} {\bibfnamefont {A.}~\bibnamefont {Ac{\'i}n}},
  \bibinfo {author} {\bibfnamefont {S.}~\bibnamefont {Massar}}, \bibinfo
  {author} {\bibfnamefont {A.~B.}\ \bibnamefont {de~La~Giroday}}, \bibinfo
  {author} {\bibfnamefont {D.~N.}\ \bibnamefont {Matsukevich}}, \bibinfo
  {author} {\bibfnamefont {P.}~\bibnamefont {Maunz}}, \bibinfo {author}
  {\bibfnamefont {S.}~\bibnamefont {Olmschenk}}, \bibinfo {author}
  {\bibfnamefont {D.}~\bibnamefont {Hayes}}, \bibinfo {author} {\bibfnamefont
  {L.}~\bibnamefont {Luo}}, \bibinfo {author} {\bibfnamefont {T.~A.}\
  \bibnamefont {Manning}},  \emph {et~al.},\ }\bibfield  {title} {\enquote
  {\bibinfo {title} {Random numbers certified by bell's theorem},}\ }\href
  {\doibase 10.1038/nature09008} {\bibfield  {journal} {\bibinfo  {journal}
  {Nature}\ }\textbf {\bibinfo {volume} {464}},\ \bibinfo {pages} {1021}
  (\bibinfo {year} {2010})}\BibitemShut {NoStop}%
\bibitem [{\citenamefont {Deng}\ and\ \citenamefont
  {Duan}(2013)}]{Deng2013Fault-tolerant}%
  \BibitemOpen
  \bibfield  {author} {\bibinfo {author} {\bibfnamefont {D.-L.}\ \bibnamefont
  {Deng}}\ and\ \bibinfo {author} {\bibfnamefont {L.-M.}\ \bibnamefont
  {Duan}},\ }\bibfield  {title} {\enquote {\bibinfo {title} {Fault-tolerant
  quantum random-number generator certified by majorana fermions},}\ }\href
  {\doibase 10.1103/PhysRevA.88.012323} {\bibfield  {journal} {\bibinfo
  {journal} {Phys. Rev. A}\ }\textbf {\bibinfo {volume} {88}},\ \bibinfo
  {pages} {012323} (\bibinfo {year} {2013})}\BibitemShut {NoStop}%
\bibitem [{\citenamefont {Herrero-Collantes}\ and\ \citenamefont
  {Garcia-Escartin}(2017)}]{Herrero2017Quantum}%
  \BibitemOpen
  \bibfield  {author} {\bibinfo {author} {\bibfnamefont {M.}~\bibnamefont
  {Herrero-Collantes}}\ and\ \bibinfo {author} {\bibfnamefont {J.~C.}\
  \bibnamefont {Garcia-Escartin}},\ }\bibfield  {title} {\enquote {\bibinfo
  {title} {Quantum random number generators},}\ }\href {\doibase
  10.1103/RevModPhys.89.015004} {\bibfield  {journal} {\bibinfo  {journal}
  {Rev. Mod. Phys.}\ }\textbf {\bibinfo {volume} {89}},\ \bibinfo {pages}
  {015004} (\bibinfo {year} {2017})}\BibitemShut {NoStop}%
\bibitem [{\citenamefont {Miller}\ and\ \citenamefont
  {Shi}(2014)}]{Miller2014Robust}%
  \BibitemOpen
  \bibfield  {author} {\bibinfo {author} {\bibfnamefont {C.~A.}\ \bibnamefont
  {Miller}}\ and\ \bibinfo {author} {\bibfnamefont {Y.}~\bibnamefont {Shi}},\
  }\bibfield  {title} {\enquote {\bibinfo {title} {Robust protocols for
  securely expanding randomness and distributing keys using untrusted quantum
  devices},}\ }in\ \href@noop {} {\emph {\bibinfo {booktitle} {Proceedings of
  the 46th Annual ACM Symposium on Theory of Computing}}}\ (\bibinfo
  {organization} {ACM},\ \bibinfo {year} {2014})\ pp.\ \bibinfo {pages}
  {417--426}\BibitemShut {NoStop}%
\bibitem [{\citenamefont {Michalski}\ \emph {et~al.}(2013)\citenamefont
  {Michalski}, \citenamefont {Carbonell},\ and\ \citenamefont
  {Mitchell}}]{Michalski2013Machine}%
  \BibitemOpen
  \bibfield  {author} {\bibinfo {author} {\bibfnamefont {R.~S.}\ \bibnamefont
  {Michalski}}, \bibinfo {author} {\bibfnamefont {J.~G.}\ \bibnamefont
  {Carbonell}}, \ and\ \bibinfo {author} {\bibfnamefont {T.~M.}\ \bibnamefont
  {Mitchell}},\ }\href@noop {} {\emph {\bibinfo {title} {Machine learning: An
  artificial intelligence approach}}}\ (\bibinfo  {publisher} {Springer Science
  \& Business Media},\ \bibinfo {year} {2013})\BibitemShut {NoStop}%
\bibitem [{\citenamefont {Jordan}\ and\ \citenamefont
  {Mitchell}(2015)}]{Jordan2015Machine}%
  \BibitemOpen
  \bibfield  {author} {\bibinfo {author} {\bibfnamefont {M.}~\bibnamefont
  {Jordan}}\ and\ \bibinfo {author} {\bibfnamefont {T.}~\bibnamefont
  {Mitchell}},\ }\bibfield  {title} {\enquote {\bibinfo {title} {Machine
  learning: Trends, perspectives, and prospects},}\ }\href {\doibase
  10.1126/science.aaa8415} {\bibfield  {journal} {\bibinfo  {journal}
  {Science}\ }\textbf {\bibinfo {volume} {349}},\ \bibinfo {pages} {255}
  (\bibinfo {year} {2015})}\BibitemShut {NoStop}%
\bibitem [{\citenamefont {LeCun}\ \emph {et~al.}(2015)\citenamefont {LeCun},
  \citenamefont {Bengio},\ and\ \citenamefont {Hinton}}]{Lecun2015Deep}%
  \BibitemOpen
  \bibfield  {author} {\bibinfo {author} {\bibfnamefont {Y.}~\bibnamefont
  {LeCun}}, \bibinfo {author} {\bibfnamefont {Y.}~\bibnamefont {Bengio}}, \
  and\ \bibinfo {author} {\bibfnamefont {G.}~\bibnamefont {Hinton}},\
  }\bibfield  {title} {\enquote {\bibinfo {title} {Deep learning},}\ }\href
  {\doibase 10.1038/nature14539} {\bibfield  {journal} {\bibinfo  {journal}
  {Nature}\ }\textbf {\bibinfo {volume} {521}},\ \bibinfo {pages} {436}
  (\bibinfo {year} {2015})}\BibitemShut {NoStop}%
\bibitem [{\citenamefont {Amico}\ \emph {et~al.}(2008)\citenamefont {Amico},
  \citenamefont {Fazio}, \citenamefont {Osterloh},\ and\ \citenamefont
  {Vedral}}]{Amico2008Entanglement}%
  \BibitemOpen
  \bibfield  {author} {\bibinfo {author} {\bibfnamefont {L.}~\bibnamefont
  {Amico}}, \bibinfo {author} {\bibfnamefont {R.}~\bibnamefont {Fazio}},
  \bibinfo {author} {\bibfnamefont {A.}~\bibnamefont {Osterloh}}, \ and\
  \bibinfo {author} {\bibfnamefont {V.}~\bibnamefont {Vedral}},\ }\bibfield
  {title} {\enquote {\bibinfo {title} {Entanglement in many-body systems},}\
  }\href {\doibase 10.1103/RevModPhys.80.517} {\bibfield  {journal} {\bibinfo
  {journal} {Rev. Mod. Phys.}\ }\textbf {\bibinfo {volume} {80}},\ \bibinfo
  {pages} {517} (\bibinfo {year} {2008})}\BibitemShut {NoStop}%
\bibitem [{\citenamefont {Babai}\ \emph {et~al.}(1991)\citenamefont {Babai},
  \citenamefont {Fortnow},\ and\ \citenamefont {Lund}}]{Babai1991Non}%
  \BibitemOpen
  \bibfield  {author} {\bibinfo {author} {\bibfnamefont {L.}~\bibnamefont
  {Babai}}, \bibinfo {author} {\bibfnamefont {L.}~\bibnamefont {Fortnow}}, \
  and\ \bibinfo {author} {\bibfnamefont {C.}~\bibnamefont {Lund}},\ }\bibfield
  {title} {\enquote {\bibinfo {title} {Non-deterministic exponential time has
  two-prover interactive protocols},}\ }\href
  {https://link.springer.com/article/10.1007%2FBF01200056?LI=true} {\bibfield
  {journal} {\bibinfo  {journal} {Comput. Complex.}\ }\textbf {\bibinfo
  {volume} {1}},\ \bibinfo {pages} {3} (\bibinfo {year} {1991})}\BibitemShut
  {NoStop}%
\bibitem [{\citenamefont {Tura}\ \emph {et~al.}(2014)\citenamefont {Tura},
  \citenamefont {Augusiak}, \citenamefont {Sainz}, \citenamefont {V{\'e}rtesi},
  \citenamefont {Lewenstein},\ and\ \citenamefont
  {Ac{\'\i}n}}]{Tura2014Detecting}%
  \BibitemOpen
  \bibfield  {author} {\bibinfo {author} {\bibfnamefont {J.}~\bibnamefont
  {Tura}}, \bibinfo {author} {\bibfnamefont {R.}~\bibnamefont {Augusiak}},
  \bibinfo {author} {\bibfnamefont {A.~B.}\ \bibnamefont {Sainz}}, \bibinfo
  {author} {\bibfnamefont {T.}~\bibnamefont {V{\'e}rtesi}}, \bibinfo {author}
  {\bibfnamefont {M.}~\bibnamefont {Lewenstein}}, \ and\ \bibinfo {author}
  {\bibfnamefont {A.}~\bibnamefont {Ac{\'\i}n}},\ }\bibfield  {title} {\enquote
  {\bibinfo {title} {Detecting nonlocality in many-body quantum states},}\
  }\href {\doibase 10.1126/science.1247715} {\bibfield  {journal} {\bibinfo
  {journal} {Science}\ }\textbf {\bibinfo {volume} {344}},\ \bibinfo {pages}
  {1256} (\bibinfo {year} {2014})}\BibitemShut {NoStop}%
\bibitem [{\citenamefont {Tura}\ \emph {et~al.}(2015)\citenamefont {Tura},
  \citenamefont {Augusiak}, \citenamefont {Sainz}, \citenamefont {L{\"u}cke},
  \citenamefont {Klempt}, \citenamefont {Lewenstein},\ and\ \citenamefont
  {Ac{\'\i}n}}]{Tura2015Nonlocality}%
  \BibitemOpen
  \bibfield  {author} {\bibinfo {author} {\bibfnamefont {J.}~\bibnamefont
  {Tura}}, \bibinfo {author} {\bibfnamefont {R.}~\bibnamefont {Augusiak}},
  \bibinfo {author} {\bibfnamefont {A.~B.}\ \bibnamefont {Sainz}}, \bibinfo
  {author} {\bibfnamefont {B.}~\bibnamefont {L{\"u}cke}}, \bibinfo {author}
  {\bibfnamefont {C.}~\bibnamefont {Klempt}}, \bibinfo {author} {\bibfnamefont
  {M.}~\bibnamefont {Lewenstein}}, \ and\ \bibinfo {author} {\bibfnamefont
  {A.}~\bibnamefont {Ac{\'\i}n}},\ }\bibfield  {title} {\enquote {\bibinfo
  {title} {Nonlocality in many-body quantum systems detected with two-body
  correlators},}\ }\href
  {http://www.sciencedirect.com/science/article/pii/S0003491615002869}
  {\bibfield  {journal} {\bibinfo  {journal} {Annals of Physics}\ }\textbf
  {\bibinfo {volume} {362}},\ \bibinfo {pages} {370} (\bibinfo {year}
  {2015})}\BibitemShut {NoStop}%
\bibitem [{\citenamefont {Tura}\ \emph {et~al.}(2017)\citenamefont {Tura},
  \citenamefont {De~las Cuevas}, \citenamefont {Augusiak}, \citenamefont
  {Lewenstein}, \citenamefont {Ac\'{\i}n},\ and\ \citenamefont
  {Cirac}}]{Tura2017Energy}%
  \BibitemOpen
  \bibfield  {author} {\bibinfo {author} {\bibfnamefont {J.}~\bibnamefont
  {Tura}}, \bibinfo {author} {\bibfnamefont {G.}~\bibnamefont {De~las Cuevas}},
  \bibinfo {author} {\bibfnamefont {R.}~\bibnamefont {Augusiak}}, \bibinfo
  {author} {\bibfnamefont {M.}~\bibnamefont {Lewenstein}}, \bibinfo {author}
  {\bibfnamefont {A.}~\bibnamefont {Ac\'{\i}n}}, \ and\ \bibinfo {author}
  {\bibfnamefont {J.~I.}\ \bibnamefont {Cirac}},\ }\bibfield  {title} {\enquote
  {\bibinfo {title} {Energy as a detector of nonlocality of many-body spin
  systems},}\ }\href {\doibase 10.1103/PhysRevX.7.021005} {\bibfield  {journal}
  {\bibinfo  {journal} {Phys. Rev. X}\ }\textbf {\bibinfo {volume} {7}},\
  \bibinfo {pages} {021005} (\bibinfo {year} {2017})}\BibitemShut {NoStop}%
\bibitem [{\citenamefont {Wang}\ \emph {et~al.}(2017)\citenamefont {Wang},
  \citenamefont {Singh},\ and\ \citenamefont
  {Navascu\'es}}]{Wang2017Entanglement}%
  \BibitemOpen
  \bibfield  {author} {\bibinfo {author} {\bibfnamefont {Z.}~\bibnamefont
  {Wang}}, \bibinfo {author} {\bibfnamefont {S.}~\bibnamefont {Singh}}, \ and\
  \bibinfo {author} {\bibfnamefont {M.}~\bibnamefont {Navascu\'es}},\
  }\bibfield  {title} {\enquote {\bibinfo {title} {Entanglement and nonlocality
  in infinite 1d systems},}\ }\href {\doibase 10.1103/PhysRevLett.118.230401}
  {\bibfield  {journal} {\bibinfo  {journal} {Phys. Rev. Lett.}\ }\textbf
  {\bibinfo {volume} {118}},\ \bibinfo {pages} {230401} (\bibinfo {year}
  {2017})}\BibitemShut {NoStop}%
\bibitem [{\citenamefont {Wagner}\ \emph {et~al.}(2017)\citenamefont {Wagner},
  \citenamefont {Schmied}, \citenamefont {Fadel}, \citenamefont {Treutlein},
  \citenamefont {Sangouard},\ and\ \citenamefont {Bancal}}]{Wagner2017Bell}%
  \BibitemOpen
  \bibfield  {author} {\bibinfo {author} {\bibfnamefont {S.}~\bibnamefont
  {Wagner}}, \bibinfo {author} {\bibfnamefont {R.}~\bibnamefont {Schmied}},
  \bibinfo {author} {\bibfnamefont {M.}~\bibnamefont {Fadel}}, \bibinfo
  {author} {\bibfnamefont {P.}~\bibnamefont {Treutlein}}, \bibinfo {author}
  {\bibfnamefont {N.}~\bibnamefont {Sangouard}}, \ and\ \bibinfo {author}
  {\bibfnamefont {J.-D.}\ \bibnamefont {Bancal}},\ }\bibfield  {title}
  {\enquote {\bibinfo {title} {Bell correlations in a many-body system with
  finite statistics},}\ }\href {https://arxiv.org/abs/1702.03088} {\bibfield
  {journal} {\bibinfo  {journal} {arXiv:1702.03088}\ } (\bibinfo {year}
  {2017})}\BibitemShut {NoStop}%
\bibitem [{\citenamefont {Schmied}\ \emph {et~al.}(2016)\citenamefont
  {Schmied}, \citenamefont {Bancal}, \citenamefont {Allard}, \citenamefont
  {Fadel}, \citenamefont {Scarani}, \citenamefont {Treutlein},\ and\
  \citenamefont {Sangouard}}]{Schmied2016Bell}%
  \BibitemOpen
  \bibfield  {author} {\bibinfo {author} {\bibfnamefont {R.}~\bibnamefont
  {Schmied}}, \bibinfo {author} {\bibfnamefont {J.-D.}\ \bibnamefont {Bancal}},
  \bibinfo {author} {\bibfnamefont {B.}~\bibnamefont {Allard}}, \bibinfo
  {author} {\bibfnamefont {M.}~\bibnamefont {Fadel}}, \bibinfo {author}
  {\bibfnamefont {V.}~\bibnamefont {Scarani}}, \bibinfo {author} {\bibfnamefont
  {P.}~\bibnamefont {Treutlein}}, \ and\ \bibinfo {author} {\bibfnamefont
  {N.}~\bibnamefont {Sangouard}},\ }\bibfield  {title} {\enquote {\bibinfo
  {title} {Bell correlations in a bose-einstein condensate},}\ }\href {\doibase
  10.1126/science.aad8665} {\bibfield  {journal} {\bibinfo  {journal}
  {Science}\ }\textbf {\bibinfo {volume} {352}},\ \bibinfo {pages} {441}
  (\bibinfo {year} {2016})}\BibitemShut {NoStop}%
\bibitem [{\citenamefont {Batle}\ \emph {et~al.}(2016)\citenamefont {Batle},
  \citenamefont {Ooi}, \citenamefont {Abdalla},\ and\ \citenamefont
  {Bagdasaryan}}]{Batle2016Computing}%
  \BibitemOpen
  \bibfield  {author} {\bibinfo {author} {\bibfnamefont {J.}~\bibnamefont
  {Batle}}, \bibinfo {author} {\bibfnamefont {C.~R.}\ \bibnamefont {Ooi}},
  \bibinfo {author} {\bibfnamefont {S.}~\bibnamefont {Abdalla}}, \ and\
  \bibinfo {author} {\bibfnamefont {A.}~\bibnamefont {Bagdasaryan}},\
  }\bibfield  {title} {\enquote {\bibinfo {title} {Computing the maximum
  violation of a bell inequality is an np-problem},}\ }\href
  {https://link.springer.com/article/10.1007/s11128-016-1275-2} {\bibfield
  {journal} {\bibinfo  {journal} {Quantum Information Processing}\ }\textbf
  {\bibinfo {volume} {15}},\ \bibinfo {pages} {2649} (\bibinfo {year}
  {2016})}\BibitemShut {NoStop}%
\bibitem [{\citenamefont {Carleo}\ and\ \citenamefont
  {Troyer}(2017)}]{Carleo2016Solving}%
  \BibitemOpen
  \bibfield  {author} {\bibinfo {author} {\bibfnamefont {G.}~\bibnamefont
  {Carleo}}\ and\ \bibinfo {author} {\bibfnamefont {M.}~\bibnamefont
  {Troyer}},\ }\bibfield  {title} {\enquote {\bibinfo {title} {Solving the
  quantum many-body problem with artificial neural networks},}\ }\href
  {\doibase 10.1126/science.aag2302} {\bibfield  {journal} {\bibinfo  {journal}
  {Science}\ }\textbf {\bibinfo {volume} {355}},\ \bibinfo {pages} {602}
  (\bibinfo {year} {2017})}\BibitemShut {NoStop}%
\bibitem [{\citenamefont {Arsenault}\ \emph {et~al.}(2015)\citenamefont
  {Arsenault}, \citenamefont {von Lilienfeld},\ and\ \citenamefont
  {Millis}}]{Arsenault2015Machine}%
  \BibitemOpen
  \bibfield  {author} {\bibinfo {author} {\bibfnamefont {L.-F.}\ \bibnamefont
  {Arsenault}}, \bibinfo {author} {\bibfnamefont {O.~A.}\ \bibnamefont {von
  Lilienfeld}}, \ and\ \bibinfo {author} {\bibfnamefont {A.~J.}\ \bibnamefont
  {Millis}},\ }\bibfield  {title} {\enquote {\bibinfo {title} {Machine learning
  for many-body physics: efficient solution of dynamical mean-field theory},}\
  }\href {https://arxiv.org/abs/1506.08858} {\bibfield  {journal} {\bibinfo
  {journal} {arXiv:1506.08858}\ } (\bibinfo {year} {2015})}\BibitemShut
  {NoStop}%
\bibitem [{\citenamefont {Zhang}\ and\ \citenamefont
  {Kim}(2017)}]{Zhang2016Triangular}%
  \BibitemOpen
  \bibfield  {author} {\bibinfo {author} {\bibfnamefont {Y.}~\bibnamefont
  {Zhang}}\ and\ \bibinfo {author} {\bibfnamefont {E.-A.}\ \bibnamefont
  {Kim}},\ }\bibfield  {title} {\enquote {\bibinfo {title} {Quantum loop
  topography for machine learning},}\ }\href {\doibase
  10.1103/PhysRevLett.118.216401} {\bibfield  {journal} {\bibinfo  {journal}
  {Phys. Rev. Lett.}\ }\textbf {\bibinfo {volume} {118}},\ \bibinfo {pages}
  {216401} (\bibinfo {year} {2017})}\BibitemShut {NoStop}%
\bibitem [{\citenamefont {Carrasquilla}\ and\ \citenamefont
  {Melko}(2017)}]{Carrasquilla2017Machine}%
  \BibitemOpen
  \bibfield  {author} {\bibinfo {author} {\bibfnamefont {J.}~\bibnamefont
  {Carrasquilla}}\ and\ \bibinfo {author} {\bibfnamefont {R.~G.}\ \bibnamefont
  {Melko}},\ }\bibfield  {title} {\enquote {\bibinfo {title} {Machine learning
  phases of matter},}\ }\href {\doibase 10.1038/nphys4035} {\bibfield
  {journal} {\bibinfo  {journal} {Nat. Phys.}\ }\textbf {\bibinfo {volume}
  {13}},\ \bibinfo {pages} {431} (\bibinfo {year} {2017})}\BibitemShut
  {NoStop}%
\bibitem [{\citenamefont {van Nieuwenburg}\ \emph {et~al.}(2017)\citenamefont
  {van Nieuwenburg}, \citenamefont {Liu},\ and\ \citenamefont
  {Huber}}]{van2017Learning}%
  \BibitemOpen
  \bibfield  {author} {\bibinfo {author} {\bibfnamefont {E.~P.}\ \bibnamefont
  {van Nieuwenburg}}, \bibinfo {author} {\bibfnamefont {Y.-H.}\ \bibnamefont
  {Liu}}, \ and\ \bibinfo {author} {\bibfnamefont {S.~D.}\ \bibnamefont
  {Huber}},\ }\bibfield  {title} {\enquote {\bibinfo {title} {Learning phase
  transitions by confusion},}\ }\href {\doibase 10.1038/nphys4037} {\bibfield
  {journal} {\bibinfo  {journal} {Nature Physics}\ }\textbf {\bibinfo {volume}
  {13}},\ \bibinfo {pages} {435} (\bibinfo {year} {2017})}\BibitemShut
  {NoStop}%
\bibitem [{\citenamefont {Deng}\ \emph {et~al.}(2016)\citenamefont {Deng},
  \citenamefont {Li},\ and\ \citenamefont {Sarma}}]{Deng2016Exact}%
  \BibitemOpen
  \bibfield  {author} {\bibinfo {author} {\bibfnamefont {D.-L.}\ \bibnamefont
  {Deng}}, \bibinfo {author} {\bibfnamefont {X.}~\bibnamefont {Li}}, \ and\
  \bibinfo {author} {\bibfnamefont {S.~D.}\ \bibnamefont {Sarma}},\ }\bibfield
  {title} {\enquote {\bibinfo {title} {Exact machine learning topological
  states},}\ }\href {https://arxiv.org/abs/1609.09060} {\bibfield  {journal}
  {\bibinfo  {journal} {arXiv:1609.09060}\ } (\bibinfo {year}
  {2016})}\BibitemShut {NoStop}%
\bibitem [{\citenamefont {Wang}(2016)}]{Wang2016Discovering}%
  \BibitemOpen
  \bibfield  {author} {\bibinfo {author} {\bibfnamefont {L.}~\bibnamefont
  {Wang}},\ }\bibfield  {title} {\enquote {\bibinfo {title} {Discovering phase
  transitions with unsupervised learning},}\ }\href {\doibase
  10.1103/PhysRevB.94.195105} {\bibfield  {journal} {\bibinfo  {journal} {Phys.
  Rev. B}\ }\textbf {\bibinfo {volume} {94}},\ \bibinfo {pages} {195105}
  (\bibinfo {year} {2016})}\BibitemShut {NoStop}%
\bibitem [{\citenamefont {Broecker}\ \emph
  {et~al.}(2017{\natexlab{a}})\citenamefont {Broecker}, \citenamefont
  {Carrasquilla}, \citenamefont {Melko},\ and\ \citenamefont
  {Trebst}}]{Broecker2017Machine}%
  \BibitemOpen
  \bibfield  {author} {\bibinfo {author} {\bibfnamefont {P.}~\bibnamefont
  {Broecker}}, \bibinfo {author} {\bibfnamefont {J.}~\bibnamefont
  {Carrasquilla}}, \bibinfo {author} {\bibfnamefont {R.~G.}\ \bibnamefont
  {Melko}}, \ and\ \bibinfo {author} {\bibfnamefont {S.}~\bibnamefont
  {Trebst}},\ }\bibfield  {title} {\enquote {\bibinfo {title} {Machine learning
  quantum phases of matter beyond the fermion sign problem},}\ }\href {\doibase
  10.1038/s41598-017-09098-0} {\bibfield  {journal} {\bibinfo  {journal} {Sci.
  Rep.}\ }\textbf {\bibinfo {volume} {7}} (\bibinfo {year}
  {2017}{\natexlab{a}}),\ 10.1038/s41598-017-09098-0}\BibitemShut {NoStop}%
\bibitem [{\citenamefont {Ch'ng}\ \emph {et~al.}(2017)\citenamefont {Ch'ng},
  \citenamefont {Carrasquilla}, \citenamefont {Melko},\ and\ \citenamefont
  {Khatami}}]{Chng2017Machine}%
  \BibitemOpen
  \bibfield  {author} {\bibinfo {author} {\bibfnamefont {K.}~\bibnamefont
  {Ch'ng}}, \bibinfo {author} {\bibfnamefont {J.}~\bibnamefont {Carrasquilla}},
  \bibinfo {author} {\bibfnamefont {R.~G.}\ \bibnamefont {Melko}}, \ and\
  \bibinfo {author} {\bibfnamefont {E.}~\bibnamefont {Khatami}},\ }\bibfield
  {title} {\enquote {\bibinfo {title} {Machine learning phases of strongly
  correlated fermions},}\ }\href {\doibase 10.1103/PhysRevX.7.031038}
  {\bibfield  {journal} {\bibinfo  {journal} {Phys. Rev. X}\ }\textbf {\bibinfo
  {volume} {7}},\ \bibinfo {pages} {031038} (\bibinfo {year}
  {2017})}\BibitemShut {NoStop}%
\bibitem [{\citenamefont {Zhang}\ \emph {et~al.}()\citenamefont {Zhang},
  \citenamefont {Melko},\ and\ \citenamefont {Kim}}]{Zhang2017Machine}%
  \BibitemOpen
  \bibfield  {author} {\bibinfo {author} {\bibfnamefont {Y.}~\bibnamefont
  {Zhang}}, \bibinfo {author} {\bibfnamefont {R.~G.}\ \bibnamefont {Melko}}, \
  and\ \bibinfo {author} {\bibfnamefont {E.-A.}\ \bibnamefont {Kim}},\
  }\bibfield  {title} {\enquote {\bibinfo {title} {Machine learning
  $\mathbb{Z}_2$ quantum spin liquids with quasi-particle statistics},}\ }\href
  {https://arxiv.org/abs/1705.01947} {\bibinfo  {journal} {arXiv:1705.01947}\
  }\BibitemShut {NoStop}%
\bibitem [{\citenamefont {Wetzel}(2017)}]{Wetzel2017Unsupervised}%
  \BibitemOpen
\bibfield  {journal} {  }\bibfield  {author} {\bibinfo {author} {\bibfnamefont
  {S.~J.}\ \bibnamefont {Wetzel}},\ }\bibfield  {title} {\enquote {\bibinfo
  {title} {Unsupervised learning of phase transitions: From principal component
  analysis to variational autoencoders},}\ }\href {\doibase
  10.1103/PhysRevE.96.022140} {\bibfield  {journal} {\bibinfo  {journal} {Phys.
  Rev. E}\ }\textbf {\bibinfo {volume} {96}},\ \bibinfo {pages} {022140}
  (\bibinfo {year} {2017})}\BibitemShut {NoStop}%
\bibitem [{\citenamefont {Hu}\ \emph {et~al.}(2017)\citenamefont {Hu},
  \citenamefont {Singh},\ and\ \citenamefont {Scalettar}}]{Hu2017Discovering}%
  \BibitemOpen
  \bibfield  {author} {\bibinfo {author} {\bibfnamefont {W.}~\bibnamefont
  {Hu}}, \bibinfo {author} {\bibfnamefont {R.~R.~P.}\ \bibnamefont {Singh}}, \
  and\ \bibinfo {author} {\bibfnamefont {R.~T.}\ \bibnamefont {Scalettar}},\
  }\bibfield  {title} {\enquote {\bibinfo {title} {Discovering phases, phase
  transitions, and crossovers through unsupervised machine learning: A critical
  examination},}\ }\href {\doibase 10.1103/PhysRevE.95.062122} {\bibfield
  {journal} {\bibinfo  {journal} {Phys. Rev. E}\ }\textbf {\bibinfo {volume}
  {95}},\ \bibinfo {pages} {062122} (\bibinfo {year} {2017})}\BibitemShut
  {NoStop}%
\bibitem [{\citenamefont {Yoshioka}\ \emph {et~al.}(2017)\citenamefont
  {Yoshioka}, \citenamefont {Akagi},\ and\ \citenamefont
  {Katsura}}]{Yoshioka2017Learning}%
  \BibitemOpen
  \bibfield  {author} {\bibinfo {author} {\bibfnamefont {N.}~\bibnamefont
  {Yoshioka}}, \bibinfo {author} {\bibfnamefont {Y.}~\bibnamefont {Akagi}}, \
  and\ \bibinfo {author} {\bibfnamefont {H.}~\bibnamefont {Katsura}},\
  }\bibfield  {title} {\enquote {\bibinfo {title} {Learning disordered
  topological phases by statistical recovery of symmetry},}\ }\href
  {https://arxiv.org/abs/1709.05790} {\bibfield  {journal} {\bibinfo  {journal}
  {arXiv:1709.05790}\ } (\bibinfo {year} {2017})}\BibitemShut {NoStop}%
\bibitem [{\citenamefont {Torlai}\ and\ \citenamefont
  {Melko}(2016)}]{Torlai2016Learning}%
  \BibitemOpen
  \bibfield  {author} {\bibinfo {author} {\bibfnamefont {G.}~\bibnamefont
  {Torlai}}\ and\ \bibinfo {author} {\bibfnamefont {R.~G.}\ \bibnamefont
  {Melko}},\ }\bibfield  {title} {\enquote {\bibinfo {title} {Learning
  thermodynamics with boltzmann machines},}\ }\href {\doibase
  10.1103/PhysRevB.94.165134} {\bibfield  {journal} {\bibinfo  {journal} {Phys.
  Rev. B}\ }\textbf {\bibinfo {volume} {94}},\ \bibinfo {pages} {165134}
  (\bibinfo {year} {2016})}\BibitemShut {NoStop}%
\bibitem [{\citenamefont {Aoki}\ and\ \citenamefont
  {Kobayashi}(2016)}]{Aoki2016restricted}%
  \BibitemOpen
  \bibfield  {author} {\bibinfo {author} {\bibfnamefont {K.-I.}\ \bibnamefont
  {Aoki}}\ and\ \bibinfo {author} {\bibfnamefont {T.}~\bibnamefont
  {Kobayashi}},\ }\bibfield  {title} {\enquote {\bibinfo {title} {Restricted
  boltzmann machines for the long range ising models},}\ }\href
  {http://www.worldscientific.com/doi/abs/10.1142/S0217984916504017} {\bibfield
   {journal} {\bibinfo  {journal} {Mod. Phys. Lett. B}\ ,\ \bibinfo {pages}
  {1650401}} (\bibinfo {year} {2016})}\BibitemShut {NoStop}%
\bibitem [{\citenamefont {You}\ \emph {et~al.}(2017)\citenamefont {You},
  \citenamefont {Yang},\ and\ \citenamefont {Qi}}]{You2017Machine}%
  \BibitemOpen
  \bibfield  {author} {\bibinfo {author} {\bibfnamefont {Y.-Z.}\ \bibnamefont
  {You}}, \bibinfo {author} {\bibfnamefont {Z.}~\bibnamefont {Yang}}, \ and\
  \bibinfo {author} {\bibfnamefont {X.-L.}\ \bibnamefont {Qi}},\ }\bibfield
  {title} {\enquote {\bibinfo {title} {Machine learning spatial geometry from
  entanglement features},}\ }\href {https://arxiv.org/abs/1709.01223}
  {\bibfield  {journal} {\bibinfo  {journal} {arXiv:1709.01223}\ } (\bibinfo
  {year} {2017})}\BibitemShut {NoStop}%
\bibitem [{\citenamefont {Torlai}\ \emph {et~al.}(2017)\citenamefont {Torlai},
  \citenamefont {Mazzola}, \citenamefont {Carrasquilla}, \citenamefont
  {Troyer}, \citenamefont {Melko},\ and\ \citenamefont
  {Carleo}}]{Torlai2017Many}%
  \BibitemOpen
  \bibfield  {author} {\bibinfo {author} {\bibfnamefont {G.}~\bibnamefont
  {Torlai}}, \bibinfo {author} {\bibfnamefont {G.}~\bibnamefont {Mazzola}},
  \bibinfo {author} {\bibfnamefont {J.}~\bibnamefont {Carrasquilla}}, \bibinfo
  {author} {\bibfnamefont {M.}~\bibnamefont {Troyer}}, \bibinfo {author}
  {\bibfnamefont {R.}~\bibnamefont {Melko}}, \ and\ \bibinfo {author}
  {\bibfnamefont {G.}~\bibnamefont {Carleo}},\ }\bibfield  {title} {\enquote
  {\bibinfo {title} {Many-body quantum state tomography with neural
  networks},}\ }\href {https://arxiv.org/abs/1703.05334} {\bibfield  {journal}
  {\bibinfo  {journal} {arXiv:1703.05334}\ } (\bibinfo {year}
  {2017})}\BibitemShut {NoStop}%
\bibitem [{\citenamefont {Pasquato}(2016)}]{Pasquato2016Detecting}%
  \BibitemOpen
  \bibfield  {author} {\bibinfo {author} {\bibfnamefont {M.}~\bibnamefont
  {Pasquato}},\ }\bibfield  {title} {\enquote {\bibinfo {title} {Detecting
  intermediate mass black holes in globular clusters with machine learning},}\
  }\href {https://arxiv.org/abs/1606.08548} {\bibfield  {journal} {\bibinfo
  {journal} {arXiv:1606.08548}\ } (\bibinfo {year} {2016})}\BibitemShut
  {NoStop}%
\bibitem [{\citenamefont {Hezaveh}\ \emph {et~al.}(2017)\citenamefont
  {Hezaveh}, \citenamefont {Perreault~Levasseur},\ and\ \citenamefont
  {Marshall}}]{Hezaveh2017Fast}%
  \BibitemOpen
  \bibfield  {author} {\bibinfo {author} {\bibfnamefont {Y.~D.}\ \bibnamefont
  {Hezaveh}}, \bibinfo {author} {\bibfnamefont {L.}~\bibnamefont
  {Perreault~Levasseur}}, \ and\ \bibinfo {author} {\bibfnamefont {P.~J.}\
  \bibnamefont {Marshall}},\ }\bibfield  {title} {\enquote {\bibinfo {title}
  {Fast automated analysis of strong gravitational lenses with convolutional
  neural networks},}\ }\href {\doibase 10.1038/nature23463} {\bibfield
  {journal} {\bibinfo  {journal} {Nature}\ }\textbf {\bibinfo {volume} {548}},\
  \bibinfo {pages} {555} (\bibinfo {year} {2017})}\BibitemShut {NoStop}%
\bibitem [{\citenamefont {Biswas}\ \emph {et~al.}(2013)\citenamefont {Biswas},
  \citenamefont {Blackburn}, \citenamefont {Cao}, \citenamefont {Essick},
  \citenamefont {Hodge}, \citenamefont {Katsavounidis}, \citenamefont {Kim},
  \citenamefont {Kim}, \citenamefont {Le~Bigot}, \citenamefont {Lee},
  \citenamefont {Oh}, \citenamefont {Oh}, \citenamefont {Son}, \citenamefont
  {Tao}, \citenamefont {Vaulin},\ and\ \citenamefont
  {Wang}}]{Rahul2013Application}%
  \BibitemOpen
  \bibfield  {author} {\bibinfo {author} {\bibfnamefont {R.}~\bibnamefont
  {Biswas}}, \bibinfo {author} {\bibfnamefont {L.}~\bibnamefont {Blackburn}},
  \bibinfo {author} {\bibfnamefont {J.}~\bibnamefont {Cao}}, \bibinfo {author}
  {\bibfnamefont {R.}~\bibnamefont {Essick}}, \bibinfo {author} {\bibfnamefont
  {K.~A.}\ \bibnamefont {Hodge}}, \bibinfo {author} {\bibfnamefont
  {E.}~\bibnamefont {Katsavounidis}}, \bibinfo {author} {\bibfnamefont
  {K.}~\bibnamefont {Kim}}, \bibinfo {author} {\bibfnamefont {Y.-M.}\
  \bibnamefont {Kim}}, \bibinfo {author} {\bibfnamefont {E.-O.}\ \bibnamefont
  {Le~Bigot}}, \bibinfo {author} {\bibfnamefont {C.-H.}\ \bibnamefont {Lee}},
  \bibinfo {author} {\bibfnamefont {J.~J.}\ \bibnamefont {Oh}}, \bibinfo
  {author} {\bibfnamefont {S.~H.}\ \bibnamefont {Oh}}, \bibinfo {author}
  {\bibfnamefont {E.~J.}\ \bibnamefont {Son}}, \bibinfo {author} {\bibfnamefont
  {Y.}~\bibnamefont {Tao}}, \bibinfo {author} {\bibfnamefont {R.}~\bibnamefont
  {Vaulin}}, \ and\ \bibinfo {author} {\bibfnamefont {X.}~\bibnamefont
  {Wang}},\ }\bibfield  {title} {\enquote {\bibinfo {title} {Application of
  machine learning algorithms to the study of noise artifacts in
  gravitational-wave data},}\ }\href {\doibase 10.1103/PhysRevD.88.062003}
  {\bibfield  {journal} {\bibinfo  {journal} {Phys. Rev. D}\ }\textbf {\bibinfo
  {volume} {88}},\ \bibinfo {pages} {062003} (\bibinfo {year}
  {2013})}\BibitemShut {NoStop}%
\bibitem [{\citenamefont {Abbott~{\it et al.}}(2016)}]{Abbott2016Observation}%
  \BibitemOpen
  \bibfield  {author} {\bibinfo {author} {\bibfnamefont {B.~P.}\ \bibnamefont
  {Abbott~{\it et al.}}} (\bibinfo {collaboration} {LIGO Scientific
  Collaboration and Virgo Collaboration}),\ }\bibfield  {title} {\enquote
  {\bibinfo {title} {Observation of gravitational waves from a binary black
  hole merger},}\ }\href {\doibase 10.1103/PhysRevLett.116.061102} {\bibfield
  {journal} {\bibinfo  {journal} {Phys. Rev. Lett.}\ }\textbf {\bibinfo
  {volume} {116}},\ \bibinfo {pages} {061102} (\bibinfo {year}
  {2016})}\BibitemShut {NoStop}%
\bibitem [{\citenamefont {Kalinin}\ \emph {et~al.}(2015)\citenamefont
  {Kalinin}, \citenamefont {Sumpter},\ and\ \citenamefont
  {Archibald}}]{Kalinin2015Big}%
  \BibitemOpen
  \bibfield  {author} {\bibinfo {author} {\bibfnamefont {S.~V.}\ \bibnamefont
  {Kalinin}}, \bibinfo {author} {\bibfnamefont {B.~G.}\ \bibnamefont
  {Sumpter}}, \ and\ \bibinfo {author} {\bibfnamefont {R.~K.}\ \bibnamefont
  {Archibald}},\ }\bibfield  {title} {\enquote {\bibinfo {title}
  {Big-deep-smart data in imaging for guiding materials design},}\ }\href
  {\doibase 10.1038/nmat4395} {\bibfield  {journal} {\bibinfo  {journal} {Nat.
  Mater.}\ }\textbf {\bibinfo {volume} {14}},\ \bibinfo {pages} {973} (\bibinfo
  {year} {2015})}\BibitemShut {NoStop}%
\bibitem [{\citenamefont {Schoenholz}\ \emph {et~al.}(2016)\citenamefont
  {Schoenholz}, \citenamefont {Cubuk}, \citenamefont {Sussman}, \citenamefont
  {Kaxiras},\ and\ \citenamefont {Liu}}]{Schoenholz2016Structural}%
  \BibitemOpen
  \bibfield  {author} {\bibinfo {author} {\bibfnamefont {S.~S.}\ \bibnamefont
  {Schoenholz}}, \bibinfo {author} {\bibfnamefont {E.~D.}\ \bibnamefont
  {Cubuk}}, \bibinfo {author} {\bibfnamefont {D.~M.}\ \bibnamefont {Sussman}},
  \bibinfo {author} {\bibfnamefont {E.}~\bibnamefont {Kaxiras}}, \ and\
  \bibinfo {author} {\bibfnamefont {A.~J.}\ \bibnamefont {Liu}},\ }\bibfield
  {title} {\enquote {\bibinfo {title} {A structural approach to relaxation in
  glassy liquids},}\ }\href {\doibase 10.1038/nphys3644} {\bibfield  {journal}
  {\bibinfo  {journal} {Nat. Phys.}\ }\textbf {\bibinfo {volume} {12}},\
  \bibinfo {pages} {469} (\bibinfo {year} {2016})}\BibitemShut {NoStop}%
\bibitem [{\citenamefont {Liu}\ \emph {et~al.}(2017)\citenamefont {Liu},
  \citenamefont {Qi}, \citenamefont {Meng},\ and\ \citenamefont
  {Fu}}]{Liu2017Self}%
  \BibitemOpen
  \bibfield  {author} {\bibinfo {author} {\bibfnamefont {J.}~\bibnamefont
  {Liu}}, \bibinfo {author} {\bibfnamefont {Y.}~\bibnamefont {Qi}}, \bibinfo
  {author} {\bibfnamefont {Z.~Y.}\ \bibnamefont {Meng}}, \ and\ \bibinfo
  {author} {\bibfnamefont {L.}~\bibnamefont {Fu}},\ }\bibfield  {title}
  {\enquote {\bibinfo {title} {Self-learning monte carlo method},}\ }\href
  {\doibase 10.1103/PhysRevB.95.041101} {\bibfield  {journal} {\bibinfo
  {journal} {Phys. Rev. B}\ }\textbf {\bibinfo {volume} {95}},\ \bibinfo
  {pages} {041101} (\bibinfo {year} {2017})}\BibitemShut {NoStop}%
\bibitem [{\citenamefont {Huang}\ and\ \citenamefont
  {Wang}(2017)}]{Huang2017Accelerated}%
  \BibitemOpen
  \bibfield  {author} {\bibinfo {author} {\bibfnamefont {L.}~\bibnamefont
  {Huang}}\ and\ \bibinfo {author} {\bibfnamefont {L.}~\bibnamefont {Wang}},\
  }\bibfield  {title} {\enquote {\bibinfo {title} {Accelerated monte carlo
  simulations with restricted boltzmann machines},}\ }\href {\doibase
  10.1103/PhysRevB.95.035105} {\bibfield  {journal} {\bibinfo  {journal} {Phys.
  Rev. B}\ }\textbf {\bibinfo {volume} {95}},\ \bibinfo {pages} {035105}
  (\bibinfo {year} {2017})}\BibitemShut {NoStop}%
\bibitem [{\citenamefont {Torlai}\ and\ \citenamefont
  {Melko}(2017)}]{Torlai2017Neural}%
  \BibitemOpen
  \bibfield  {author} {\bibinfo {author} {\bibfnamefont {G.}~\bibnamefont
  {Torlai}}\ and\ \bibinfo {author} {\bibfnamefont {R.~G.}\ \bibnamefont
  {Melko}},\ }\bibfield  {title} {\enquote {\bibinfo {title} {Neural decoder
  for topological codes},}\ }\href {\doibase 10.1103/PhysRevLett.119.030501}
  {\bibfield  {journal} {\bibinfo  {journal} {Phys. Rev. Lett.}\ }\textbf
  {\bibinfo {volume} {119}},\ \bibinfo {pages} {030501} (\bibinfo {year}
  {2017})}\BibitemShut {NoStop}%
\bibitem [{\citenamefont {Gao}\ and\ \citenamefont
  {Duan}(2017)}]{Gao2017Efficient}%
  \BibitemOpen
  \bibfield  {author} {\bibinfo {author} {\bibfnamefont {X.}~\bibnamefont
  {Gao}}\ and\ \bibinfo {author} {\bibfnamefont {L.-M.}\ \bibnamefont {Duan}},\
  }\bibfield  {title} {\enquote {\bibinfo {title} {Efficient representation of
  quantum many-body states with deep neural networks},}\ }\href {\doibase
  10.1038/s41467-017-00705-2} {\bibfield  {journal} {\bibinfo  {journal} {Nat.
  Commu.}\ ,\ \bibinfo {pages} {662}} (\bibinfo {year} {2017})}\BibitemShut
  {NoStop}%
\bibitem [{\citenamefont {Chen}\ \emph {et~al.}(2017)\citenamefont {Chen},
  \citenamefont {Cheng}, \citenamefont {Xie}, \citenamefont {Wang},\ and\
  \citenamefont {Xiang}}]{Chen2017Equivalence}%
  \BibitemOpen
  \bibfield  {author} {\bibinfo {author} {\bibfnamefont {J.}~\bibnamefont
  {Chen}}, \bibinfo {author} {\bibfnamefont {S.}~\bibnamefont {Cheng}},
  \bibinfo {author} {\bibfnamefont {H.}~\bibnamefont {Xie}}, \bibinfo {author}
  {\bibfnamefont {L.}~\bibnamefont {Wang}}, \ and\ \bibinfo {author}
  {\bibfnamefont {T.}~\bibnamefont {Xiang}},\ }\bibfield  {title} {\enquote
  {\bibinfo {title} {On the equivalence of restricted boltzmann machines and
  tensor network states},}\ }\href {https://arxiv.org/abs/1701.04831}
  {\bibfield  {journal} {\bibinfo  {journal} {arXiv: 1701.04831}\ } (\bibinfo
  {year} {2017})}\BibitemShut {NoStop}%
\bibitem [{\citenamefont {Huang}\ and\ \citenamefont
  {Moore}(2017)}]{Huang2017Neural}%
  \BibitemOpen
  \bibfield  {author} {\bibinfo {author} {\bibfnamefont {Y.}~\bibnamefont
  {Huang}}\ and\ \bibinfo {author} {\bibfnamefont {J.~E.}\ \bibnamefont
  {Moore}},\ }\bibfield  {title} {\enquote {\bibinfo {title} {Neural network
  representation of tensor network and chiral states},}\ }\href
  {https://arxiv.org/abs/1701.06246} {\bibfield  {journal} {\bibinfo  {journal}
  {arXiv:1701.06246}\ } (\bibinfo {year} {2017})}\BibitemShut {NoStop}%
\bibitem [{\citenamefont {Schindler}\ \emph {et~al.}(2017)\citenamefont
  {Schindler}, \citenamefont {Regnault},\ and\ \citenamefont
  {Neupert}}]{Schindler2017Probing}%
  \BibitemOpen
  \bibfield  {author} {\bibinfo {author} {\bibfnamefont {F.}~\bibnamefont
  {Schindler}}, \bibinfo {author} {\bibfnamefont {N.}~\bibnamefont {Regnault}},
  \ and\ \bibinfo {author} {\bibfnamefont {T.}~\bibnamefont {Neupert}},\
  }\bibfield  {title} {\enquote {\bibinfo {title} {Probing many-body
  localization with neural networks},}\ }\href {\doibase
  10.1103/PhysRevB.95.245134} {\bibfield  {journal} {\bibinfo  {journal} {Phys.
  Rev. B}\ }\textbf {\bibinfo {volume} {95}},\ \bibinfo {pages} {245134}
  (\bibinfo {year} {2017})}\BibitemShut {NoStop}%
\bibitem [{\citenamefont {Cai}(2017)}]{Cai2017Approximating}%
  \BibitemOpen
  \bibfield  {author} {\bibinfo {author} {\bibfnamefont {Z.}~\bibnamefont
  {Cai}},\ }\bibfield  {title} {\enquote {\bibinfo {title} {Approximating
  quantum many-body wave-functions using artificial neural networks},}\ }\href
  {https://arxiv.org/abs/1704.05148} {\bibfield  {journal} {\bibinfo  {journal}
  {arXiv:1704.05148}\ } (\bibinfo {year} {2017})}\BibitemShut {NoStop}%
\bibitem [{\citenamefont {Broecker}\ \emph
  {et~al.}(2017{\natexlab{b}})\citenamefont {Broecker}, \citenamefont
  {Assaad},\ and\ \citenamefont {Trebst}}]{Broecker2017Quantum}%
  \BibitemOpen
  \bibfield  {author} {\bibinfo {author} {\bibfnamefont {P.}~\bibnamefont
  {Broecker}}, \bibinfo {author} {\bibfnamefont {F.~F.}\ \bibnamefont
  {Assaad}}, \ and\ \bibinfo {author} {\bibfnamefont {S.}~\bibnamefont
  {Trebst}},\ }\bibfield  {title} {\enquote {\bibinfo {title} {Quantum phase
  recognition via unsupervised machine learning},}\ }\href
  {https://arxiv.org/abs/1707.00663} {\bibfield  {journal} {\bibinfo  {journal}
  {arXiv:1707.00663}\ } (\bibinfo {year} {2017}{\natexlab{b}})}\BibitemShut
  {NoStop}%
\bibitem [{\citenamefont {Nomura}\ \emph {et~al.}(2017)\citenamefont {Nomura},
  \citenamefont {Darmawan}, \citenamefont {Yamaji},\ and\ \citenamefont
  {Imada}}]{Nomura2017Restricted}%
  \BibitemOpen
  \bibfield  {author} {\bibinfo {author} {\bibfnamefont {Y.}~\bibnamefont
  {Nomura}}, \bibinfo {author} {\bibfnamefont {A.}~\bibnamefont {Darmawan}},
  \bibinfo {author} {\bibfnamefont {Y.}~\bibnamefont {Yamaji}}, \ and\ \bibinfo
  {author} {\bibfnamefont {M.}~\bibnamefont {Imada}},\ }\bibfield  {title}
  {\enquote {\bibinfo {title} {Restricted-boltzmann-machine learning for
  solving strongly correlated quantum systems},}\ }\href
  {https://arxiv.org/abs/1709.06475} {\bibfield  {journal} {\bibinfo  {journal}
  {arXiv:1709.06475}\ } (\bibinfo {year} {2017})}\BibitemShut {NoStop}%
\bibitem [{\citenamefont {Biamonte}\ \emph {et~al.}(2017)\citenamefont
  {Biamonte}, \citenamefont {Wittek}, \citenamefont {Pancotti}, \citenamefont
  {Rebentrost}, \citenamefont {Wiebe},\ and\ \citenamefont
  {Lloyd}}]{Biamonte2017Quantum}%
  \BibitemOpen
  \bibfield  {author} {\bibinfo {author} {\bibfnamefont {J.}~\bibnamefont
  {Biamonte}}, \bibinfo {author} {\bibfnamefont {P.}~\bibnamefont {Wittek}},
  \bibinfo {author} {\bibfnamefont {N.}~\bibnamefont {Pancotti}}, \bibinfo
  {author} {\bibfnamefont {P.}~\bibnamefont {Rebentrost}}, \bibinfo {author}
  {\bibfnamefont {N.}~\bibnamefont {Wiebe}}, \ and\ \bibinfo {author}
  {\bibfnamefont {S.}~\bibnamefont {Lloyd}},\ }\bibfield  {title} {\enquote
  {\bibinfo {title} {Quantum machine learning},}\ }\href {\doibase
  10.1038/nature23474} {\bibfield  {journal} {\bibinfo  {journal} {Nature}\ ,\
  \bibinfo {pages} {195}} (\bibinfo {year} {2017})}\BibitemShut {NoStop}%
\bibitem [{\citenamefont {Hinton}\ and\ \citenamefont
  {Salakhutdinov}(2006)}]{Hinton2006Reducing}%
  \BibitemOpen
  \bibfield  {author} {\bibinfo {author} {\bibfnamefont {G.~E.}\ \bibnamefont
  {Hinton}}\ and\ \bibinfo {author} {\bibfnamefont {R.~R.}\ \bibnamefont
  {Salakhutdinov}},\ }\bibfield  {title} {\enquote {\bibinfo {title} {Reducing
  the dimensionality of data with neural networks},}\ }\href {\doibase
  10.1126/science.1127647} {\bibfield  {journal} {\bibinfo  {journal}
  {Science}\ }\textbf {\bibinfo {volume} {313}},\ \bibinfo {pages} {504}
  (\bibinfo {year} {2006})}\BibitemShut {NoStop}%
\bibitem [{\citenamefont {Salakhutdinov}\ \emph {et~al.}(2007)\citenamefont
  {Salakhutdinov}, \citenamefont {Mnih},\ and\ \citenamefont
  {Hinton}}]{Salakhutdinov2007Restricted}%
  \BibitemOpen
  \bibfield  {author} {\bibinfo {author} {\bibfnamefont {R.}~\bibnamefont
  {Salakhutdinov}}, \bibinfo {author} {\bibfnamefont {A.}~\bibnamefont {Mnih}},
  \ and\ \bibinfo {author} {\bibfnamefont {G.}~\bibnamefont {Hinton}},\
  }\bibfield  {title} {\enquote {\bibinfo {title} {Restricted boltzmann
  machines for collaborative filtering},}\ }in\ \href
  {http://dl.acm.org/citation.cfm?id=1273596} {\emph {\bibinfo {booktitle}
  {Proceedings of the 24th international conference on Machine learning}}}\
  (\bibinfo {organization} {ACM},\ \bibinfo {year} {2007})\ pp.\ \bibinfo
  {pages} {791--798}\BibitemShut {NoStop}%
\bibitem [{\citenamefont {Larochelle}\ and\ \citenamefont
  {Bengio}(2008)}]{Larochelle2008Classification}%
  \BibitemOpen
  \bibfield  {author} {\bibinfo {author} {\bibfnamefont {H.}~\bibnamefont
  {Larochelle}}\ and\ \bibinfo {author} {\bibfnamefont {Y.}~\bibnamefont
  {Bengio}},\ }\bibfield  {title} {\enquote {\bibinfo {title} {Classification
  using discriminative restricted boltzmann machines},}\ }in\ \href
  {http://dl.acm.org/citation.cfm?id=1390224} {\emph {\bibinfo {booktitle}
  {Proceedings of the 25th international conference on Machine learning}}}\
  (\bibinfo {organization} {ACM},\ \bibinfo {year} {2008})\ pp.\ \bibinfo
  {pages} {536--543}\BibitemShut {NoStop}%
\bibitem [{\citenamefont {Kolmogorov}(1963)}]{Kolmogorov1963Representation}%
  \BibitemOpen
  \bibfield  {author} {\bibinfo {author} {\bibfnamefont {A.~N.}\ \bibnamefont
  {Kolmogorov}},\ }\bibfield  {title} {\enquote {\bibinfo {title} {On the
  representation of continuous functions of many variables by superposition of
  continuous functions of one variable and addition},}\ }\href@noop {}
  {\bibfield  {journal} {\bibinfo  {journal} {Amer. Math. Soc. Transl}\
  }\textbf {\bibinfo {volume} {28}},\ \bibinfo {pages} {55} (\bibinfo {year}
  {1963})}\BibitemShut {NoStop}%
\bibitem [{\citenamefont {Le~Roux}\ and\ \citenamefont
  {Bengio}(2008)}]{Le2008Representational}%
  \BibitemOpen
  \bibfield  {author} {\bibinfo {author} {\bibfnamefont {N.}~\bibnamefont
  {Le~Roux}}\ and\ \bibinfo {author} {\bibfnamefont {Y.}~\bibnamefont
  {Bengio}},\ }\bibfield  {title} {\enquote {\bibinfo {title} {Representational
  power of restricted boltzmann machines and deep belief networks},}\ }\href
  {http://www.mitpressjournals.org/doi/abs/10.1162/neco.2008.04-07-510#.V8uLwrXG5Bw}
  {\bibfield  {journal} {\bibinfo  {journal} {Neural Comput.}\ }\textbf
  {\bibinfo {volume} {20}},\ \bibinfo {pages} {1631} (\bibinfo {year}
  {2008})}\BibitemShut {NoStop}%
\bibitem [{\citenamefont {Hornik}(1991)}]{Hornik1991Approximation}%
  \BibitemOpen
  \bibfield  {author} {\bibinfo {author} {\bibfnamefont {K.}~\bibnamefont
  {Hornik}},\ }\bibfield  {title} {\enquote {\bibinfo {title} {Approximation
  capabilities of multilayer feedforward networks},}\ }\href
  {http://www.sciencedirect.com/science/article/pii/089360809190009T}
  {\bibfield  {journal} {\bibinfo  {journal} {Neural networks}\ }\textbf
  {\bibinfo {volume} {4}},\ \bibinfo {pages} {251} (\bibinfo {year}
  {1991})}\BibitemShut {NoStop}%
\bibitem [{\citenamefont {Deng}\ \emph {et~al.}(2017)\citenamefont {Deng},
  \citenamefont {Li},\ and\ \citenamefont {Das~Sarma}}]{Deng2017Quantum}%
  \BibitemOpen
  \bibfield  {author} {\bibinfo {author} {\bibfnamefont {D.-L.}\ \bibnamefont
  {Deng}}, \bibinfo {author} {\bibfnamefont {X.}~\bibnamefont {Li}}, \ and\
  \bibinfo {author} {\bibfnamefont {S.}~\bibnamefont {Das~Sarma}},\ }\bibfield
  {title} {\enquote {\bibinfo {title} {Quantum entanglement in neural network
  states},}\ }\href {\doibase 10.1103/PhysRevX.7.021021} {\bibfield  {journal}
  {\bibinfo  {journal} {Phys. Rev. X}\ }\textbf {\bibinfo {volume} {7}},\
  \bibinfo {pages} {021021} (\bibinfo {year} {2017})}\BibitemShut {NoStop}%
\bibitem [{\citenamefont {Einstein}\ \emph {et~al.}(1935)\citenamefont
  {Einstein}, \citenamefont {Podolsky},\ and\ \citenamefont
  {Rosen}}]{Einstein1935Can}%
  \BibitemOpen
  \bibfield  {author} {\bibinfo {author} {\bibfnamefont {A.}~\bibnamefont
  {Einstein}}, \bibinfo {author} {\bibfnamefont {B.}~\bibnamefont {Podolsky}},
  \ and\ \bibinfo {author} {\bibfnamefont {N.}~\bibnamefont {Rosen}},\
  }\bibfield  {title} {\enquote {\bibinfo {title} {Can quantum-mechanical
  description of physical reality be considered complete?}}\ }\href {\doibase
  10.1103/PhysRev.47.777} {\bibfield  {journal} {\bibinfo  {journal} {Phys.
  Rev.}\ }\textbf {\bibinfo {volume} {47}},\ \bibinfo {pages} {777} (\bibinfo
  {year} {1935})}\BibitemShut {NoStop}%
\bibitem [{\citenamefont {White}(1992)}]{White1992Density}%
  \BibitemOpen
  \bibfield  {author} {\bibinfo {author} {\bibfnamefont {S.~R.}\ \bibnamefont
  {White}},\ }\bibfield  {title} {\enquote {\bibinfo {title} {Density matrix
  formulation for quantum renormalization groups},}\ }\href {\doibase
  10.1103/PhysRevLett.69.2863} {\bibfield  {journal} {\bibinfo  {journal}
  {Phys. Rev. Lett.}\ }\textbf {\bibinfo {volume} {69}},\ \bibinfo {pages}
  {2863} (\bibinfo {year} {1992})}\BibitemShut {NoStop}%
\bibitem [{\citenamefont {Schollw{\"o}ck}(2011)}]{Schollwock2011Density}%
  \BibitemOpen
  \bibfield  {author} {\bibinfo {author} {\bibfnamefont {U.}~\bibnamefont
  {Schollw{\"o}ck}},\ }\bibfield  {title} {\enquote {\bibinfo {title} {The
  density-matrix renormalization group in the age of matrix product states},}\
  }\href {\doibase 10.1016/j.aop.2010.09.012} {\bibfield  {journal} {\bibinfo
  {journal} {Ann. of Phys.}\ }\textbf {\bibinfo {volume} {326}},\ \bibinfo
  {pages} {96} (\bibinfo {year} {2011})}\BibitemShut {NoStop}%
\bibitem [{\citenamefont {Schollw\"ock}(2005)}]{Schollwock2005TheDMRG}%
  \BibitemOpen
  \bibfield  {author} {\bibinfo {author} {\bibfnamefont {U.}~\bibnamefont
  {Schollw\"ock}},\ }\bibfield  {title} {\enquote {\bibinfo {title} {The
  density-matrix renormalization group},}\ }\href {\doibase
  10.1103/RevModPhys.77.259} {\bibfield  {journal} {\bibinfo  {journal} {Rev.
  Mod. Phys.}\ }\textbf {\bibinfo {volume} {77}},\ \bibinfo {pages} {259}
  (\bibinfo {year} {2005})}\BibitemShut {NoStop}%
\bibitem [{Bvi()}]{BvioSup}%
  \BibitemOpen
  \href@noop {} {}\bibinfo {note} {See Supplemental Material at [URL will be
  inserted by publisher] for details on the structure of the neural networks
  and the DMRG calculations, and for more numerical data.}\BibitemShut {Stop}%
\bibitem [{MLB()}]{MLBnonLa}%
  \BibitemOpen
  \href@noop {} {}\bibinfo {note} {The matrix elements of $\Lambda$ are
  specified as follows:
  $\Lambda_{0,0}=\Lambda_{0,1}=\Lambda_{1,0}=\Lambda_{2,1}=-\Lambda_{1,1}=-\Lambda_{2,0}=-\Lambda_{3,0}=-\Lambda_{3,1}=1$
  and
  $\Lambda_{0,2}=-\Lambda_{1,2}=-\Lambda_{2,2}=\Lambda_{3,2}=\Delta$}\BibitemShut
  {NoStop}%
\bibitem [{sup()}]{supplementmlb}%
  \BibitemOpen
  \href@noop {} {}\bibinfo {note} {On the even sites, we choose the
  measurements as
  $\mathcal{M}^{(2k)}_0=(\hat{\sigma}^x+\hat{\sigma}^y+\hat{\sigma}^z)/3$,
  $\mathcal{M}^{(2k)}_0=(\hat{\sigma}^x-\hat{\sigma}^y-\hat{\sigma}^z)/3$,
  $\mathcal{M}^{(2k)}_0=(-\hat{\sigma}^x+\hat{\sigma}^y-\hat{\sigma}^z)/3$, and
  $\mathcal{M}^{(2k)}_0=(-\hat{\sigma}^x-\hat{\sigma}^y+\hat{\sigma}^z)/3$;
  whereas on the odd sites, the measurement are
  $\mathcal{M}^{(2k+1)}_0=\hat{\sigma}^x$,
  $\mathcal{M}^{(2k+1)}_1=\hat{\sigma}^y$,
  $\mathcal{M}^{(2k+1)}_2=\hat{\sigma}^z$. Here, $\hat{\sigma}^x$,
  $\hat{\sigma}^y$, and $\hat{\sigma}^z$ are the usual Pauli
  matrices.}\BibitemShut {Stop}%
\bibitem [{\citenamefont {Dial}\ \emph {et~al.}(2013)\citenamefont {Dial},
  \citenamefont {Shulman}, \citenamefont {Harvey}, \citenamefont {Bluhm},
  \citenamefont {Umansky},\ and\ \citenamefont {Yacoby}}]{Dial2013Charge}%
  \BibitemOpen
  \bibfield  {author} {\bibinfo {author} {\bibfnamefont {O.~E.}\ \bibnamefont
  {Dial}}, \bibinfo {author} {\bibfnamefont {M.~D.}\ \bibnamefont {Shulman}},
  \bibinfo {author} {\bibfnamefont {S.~P.}\ \bibnamefont {Harvey}}, \bibinfo
  {author} {\bibfnamefont {H.}~\bibnamefont {Bluhm}}, \bibinfo {author}
  {\bibfnamefont {V.}~\bibnamefont {Umansky}}, \ and\ \bibinfo {author}
  {\bibfnamefont {A.}~\bibnamefont {Yacoby}},\ }\bibfield  {title} {\enquote
  {\bibinfo {title} {Charge noise spectroscopy using coherent exchange
  oscillations in a singlet-triplet qubit},}\ }\href {\doibase
  10.1103/PhysRevLett.110.146804} {\bibfield  {journal} {\bibinfo  {journal}
  {Phys. Rev. Lett.}\ }\textbf {\bibinfo {volume} {110}},\ \bibinfo {pages}
  {146804} (\bibinfo {year} {2013})}\BibitemShut {NoStop}%
\bibitem [{\citenamefont {Medford}\ \emph {et~al.}(2012)\citenamefont
  {Medford}, \citenamefont {Cywi\ifmmode~\acute{n}\else \'{n}\fi{}ski},
  \citenamefont {Barthel}, \citenamefont {Marcus}, \citenamefont {Hanson},\
  and\ \citenamefont {Gossard}}]{Medford2012Scaling}%
  \BibitemOpen
  \bibfield  {author} {\bibinfo {author} {\bibfnamefont {J.}~\bibnamefont
  {Medford}}, \bibinfo {author} {\bibfnamefont {L.}~\bibnamefont
  {Cywi\ifmmode~\acute{n}\else \'{n}\fi{}ski}}, \bibinfo {author}
  {\bibfnamefont {C.}~\bibnamefont {Barthel}}, \bibinfo {author} {\bibfnamefont
  {C.~M.}\ \bibnamefont {Marcus}}, \bibinfo {author} {\bibfnamefont {M.~P.}\
  \bibnamefont {Hanson}}, \ and\ \bibinfo {author} {\bibfnamefont {A.~C.}\
  \bibnamefont {Gossard}},\ }\bibfield  {title} {\enquote {\bibinfo {title}
  {Scaling of dynamical decoupling for spin qubits},}\ }\href {\doibase
  10.1103/PhysRevLett.108.086802} {\bibfield  {journal} {\bibinfo  {journal}
  {Phys. Rev. Lett.}\ }\textbf {\bibinfo {volume} {108}},\ \bibinfo {pages}
  {086802} (\bibinfo {year} {2012})}\BibitemShut {NoStop}%
\bibitem [{\citenamefont {Fink}\ and\ \citenamefont
  {Bluhm}(2013)}]{Fink2013Noise}%
  \BibitemOpen
  \bibfield  {author} {\bibinfo {author} {\bibfnamefont {T.}~\bibnamefont
  {Fink}}\ and\ \bibinfo {author} {\bibfnamefont {H.}~\bibnamefont {Bluhm}},\
  }\bibfield  {title} {\enquote {\bibinfo {title} {Noise spectroscopy using
  correlations of single-shot qubit readout},}\ }\href {\doibase
  10.1103/PhysRevLett.110.010403} {\bibfield  {journal} {\bibinfo  {journal}
  {Phys. Rev. Lett.}\ }\textbf {\bibinfo {volume} {110}},\ \bibinfo {pages}
  {010403} (\bibinfo {year} {2013})}\BibitemShut {NoStop}%
\bibitem [{\citenamefont {Baccari}\ \emph {et~al.}(2017)\citenamefont
  {Baccari}, \citenamefont {Cavalcanti}, \citenamefont {Wittek},\ and\
  \citenamefont {Ac\'{\i}n}}]{Baccari2017Efficient}%
  \BibitemOpen
  \bibfield  {author} {\bibinfo {author} {\bibfnamefont {F.}~\bibnamefont
  {Baccari}}, \bibinfo {author} {\bibfnamefont {D.}~\bibnamefont {Cavalcanti}},
  \bibinfo {author} {\bibfnamefont {P.}~\bibnamefont {Wittek}}, \ and\ \bibinfo
  {author} {\bibfnamefont {A.}~\bibnamefont {Ac\'{\i}n}},\ }\bibfield  {title}
  {\enquote {\bibinfo {title} {Efficient device-independent entanglement
  detection for multipartite systems},}\ }\href {\doibase
  10.1103/PhysRevX.7.021042} {\bibfield  {journal} {\bibinfo  {journal} {Phys.
  Rev. X}\ }\textbf {\bibinfo {volume} {7}},\ \bibinfo {pages} {021042}
  (\bibinfo {year} {2017})}\BibitemShut {NoStop}%
\bibitem [{\citenamefont {Masanes}(2005)}]{Masanes2005Extremal}%
  \BibitemOpen
  \bibfield  {author} {\bibinfo {author} {\bibfnamefont {L.}~\bibnamefont
  {Masanes}},\ }\bibfield  {title} {\enquote {\bibinfo {title} {Extremal
  quantum correlations for n parties with two dichotomic observables per
  site},}\ }\href {https://arxiv.org/abs/quant-ph/0512100} {\bibfield
  {journal} {\bibinfo  {journal} {arXiv: 0512100}\ } (\bibinfo {year}
  {2005})}\BibitemShut {NoStop}%
\bibitem [{\citenamefont {Toner}\ and\ \citenamefont
  {Verstraete}(2006)}]{Toner2006Monogamy}%
  \BibitemOpen
  \bibfield  {author} {\bibinfo {author} {\bibfnamefont {B.}~\bibnamefont
  {Toner}}\ and\ \bibinfo {author} {\bibfnamefont {F.}~\bibnamefont
  {Verstraete}},\ }\bibfield  {title} {\enquote {\bibinfo {title} {Monogamy of
  bell correlations and tsirelson's bound},}\ }\href
  {https://arxiv.org/abs/quant-ph/0611001} {\bibfield  {journal} {\bibinfo
  {journal} {arXiv: 0611001}\ } (\bibinfo {year} {2006})}\BibitemShut {NoStop}%
\bibitem [{\citenamefont {Clauser}\ \emph {et~al.}(1969)\citenamefont
  {Clauser}, \citenamefont {Horne}, \citenamefont {Shimony},\ and\
  \citenamefont {Holt}}]{Clauser1969Proposed}%
  \BibitemOpen
  \bibfield  {author} {\bibinfo {author} {\bibfnamefont {J.~F.}\ \bibnamefont
  {Clauser}}, \bibinfo {author} {\bibfnamefont {M.~A.}\ \bibnamefont {Horne}},
  \bibinfo {author} {\bibfnamefont {A.}~\bibnamefont {Shimony}}, \ and\
  \bibinfo {author} {\bibfnamefont {R.~A.}\ \bibnamefont {Holt}},\ }\bibfield
  {title} {\enquote {\bibinfo {title} {Proposed experiment to test local
  hidden-variable theories},}\ }\href {\doibase 10.1103/PhysRevLett.23.880}
  {\bibfield  {journal} {\bibinfo  {journal} {Phys. Rev. Lett.}\ }\textbf
  {\bibinfo {volume} {23}},\ \bibinfo {pages} {880} (\bibinfo {year}
  {1969})}\BibitemShut {NoStop}%
\bibitem [{\citenamefont {Cirel'son}(1980)}]{Cirel1980quantum}%
  \BibitemOpen
  \bibfield  {author} {\bibinfo {author} {\bibfnamefont {B.~S.}\ \bibnamefont
  {Cirel'son}},\ }\bibfield  {title} {\enquote {\bibinfo {title} {Quantum
  generalizations of bell's inequality},}\ }\href
  {https://link.springer.com/article/10.1007/BF00417500} {\bibfield  {journal}
  {\bibinfo  {journal} {Letters in Mathematical Physics}\ }\textbf {\bibinfo
  {volume} {4}},\ \bibinfo {pages} {93} (\bibinfo {year} {1980})}\BibitemShut
  {NoStop}%
\bibitem [{\citenamefont {Verstraete}\ \emph {et~al.}(2008)\citenamefont
  {Verstraete}, \citenamefont {Murg},\ and\ \citenamefont
  {Cirac}}]{Verstraete2008Matrix}%
  \BibitemOpen
  \bibfield  {author} {\bibinfo {author} {\bibfnamefont {F.}~\bibnamefont
  {Verstraete}}, \bibinfo {author} {\bibfnamefont {V.}~\bibnamefont {Murg}}, \
  and\ \bibinfo {author} {\bibfnamefont {J.~I.}\ \bibnamefont {Cirac}},\
  }\bibfield  {title} {\enquote {\bibinfo {title} {Matrix product states,
  projected entangled pair states, and variational renormalization group
  methods for quantum spin systems},}\ }\href
  {http://dx.doi.org/10.1080/14789940801912366} {\bibfield  {journal} {\bibinfo
   {journal} {Advances in Physics}\ }\textbf {\bibinfo {volume} {57}},\
  \bibinfo {pages} {143} (\bibinfo {year} {2008})}\BibitemShut {NoStop}%
\bibitem [{\citenamefont {Vidal}(2008)}]{Vidal2008Class}%
  \BibitemOpen
  \bibfield  {author} {\bibinfo {author} {\bibfnamefont {G.}~\bibnamefont
  {Vidal}},\ }\bibfield  {title} {\enquote {\bibinfo {title} {Class of quantum
  many-body states that can be efficiently simulated},}\ }\href {\doibase
  10.1103/PhysRevLett.101.110501} {\bibfield  {journal} {\bibinfo  {journal}
  {Phys. Rev. Lett.}\ }\textbf {\bibinfo {volume} {101}},\ \bibinfo {pages}
  {110501} (\bibinfo {year} {2008})}\BibitemShut {NoStop}%
\bibitem [{\citenamefont {Schmidhuber}(2015)}]{Schmidhuber2015Deep}%
  \BibitemOpen
  \bibfield  {author} {\bibinfo {author} {\bibfnamefont {J.}~\bibnamefont
  {Schmidhuber}},\ }\bibfield  {title} {\enquote {\bibinfo {title} {Deep
  learning in neural networks: An overview},}\ }\href
  {http://www.sciencedirect.com/science/article/pii/S0893608014002135}
  {\bibfield  {journal} {\bibinfo  {journal} {Neural Networks}\ }\textbf
  {\bibinfo {volume} {61}},\ \bibinfo {pages} {85} (\bibinfo {year}
  {2015})}\BibitemShut {NoStop}%
\bibitem [{\citenamefont {Bernien}\ \emph {et~al.}(2017)\citenamefont
  {Bernien}, \citenamefont {Schwartz}, \citenamefont {Keesling}, \citenamefont
  {Levine}, \citenamefont {Omran}, \citenamefont {Pichler}, \citenamefont
  {Choi}, \citenamefont {Zibrov}, \citenamefont {Endres}, \citenamefont
  {Greiner} \emph {et~al.}}]{Bernien2017Probing}%
  \BibitemOpen
  \bibfield  {author} {\bibinfo {author} {\bibfnamefont {H.}~\bibnamefont
  {Bernien}}, \bibinfo {author} {\bibfnamefont {S.}~\bibnamefont {Schwartz}},
  \bibinfo {author} {\bibfnamefont {A.}~\bibnamefont {Keesling}}, \bibinfo
  {author} {\bibfnamefont {H.}~\bibnamefont {Levine}}, \bibinfo {author}
  {\bibfnamefont {A.}~\bibnamefont {Omran}}, \bibinfo {author} {\bibfnamefont
  {H.}~\bibnamefont {Pichler}}, \bibinfo {author} {\bibfnamefont
  {S.}~\bibnamefont {Choi}}, \bibinfo {author} {\bibfnamefont {A.~S.}\
  \bibnamefont {Zibrov}}, \bibinfo {author} {\bibfnamefont {M.}~\bibnamefont
  {Endres}}, \bibinfo {author} {\bibfnamefont {M.}~\bibnamefont {Greiner}},
  \emph {et~al.},\ }\bibfield  {title} {\enquote {\bibinfo {title} {Probing
  many-body dynamics on a 51-atom quantum simulator},}\ }\href
  {https://arxiv.org/abs/1707.04344} {\bibfield  {journal} {\bibinfo  {journal}
  {arXiv:1707.04344, accepted for publication in Nature}\ } (\bibinfo {year}
  {2017})}\BibitemShut {NoStop}%
\bibitem [{\citenamefont {Zhang}\ \emph {et~al.}(2017)\citenamefont {Zhang},
  \citenamefont {Pagano}, \citenamefont {Hess}, \citenamefont {Kyprianidis},
  \citenamefont {Becker}, \citenamefont {Kaplan}, \citenamefont {Gorshkov},
  \citenamefont {Gong},\ and\ \citenamefont
  {Monroe}}]{Zhang2017ObservationDPT}%
  \BibitemOpen
  \bibfield  {author} {\bibinfo {author} {\bibfnamefont {J.}~\bibnamefont
  {Zhang}}, \bibinfo {author} {\bibfnamefont {G.}~\bibnamefont {Pagano}},
  \bibinfo {author} {\bibfnamefont {P.}~\bibnamefont {Hess}}, \bibinfo {author}
  {\bibfnamefont {A.}~\bibnamefont {Kyprianidis}}, \bibinfo {author}
  {\bibfnamefont {P.}~\bibnamefont {Becker}}, \bibinfo {author} {\bibfnamefont
  {H.}~\bibnamefont {Kaplan}}, \bibinfo {author} {\bibfnamefont
  {A.}~\bibnamefont {Gorshkov}}, \bibinfo {author} {\bibfnamefont {Z.-X.}\
  \bibnamefont {Gong}}, \ and\ \bibinfo {author} {\bibfnamefont
  {C.}~\bibnamefont {Monroe}},\ }\bibfield  {title} {\enquote {\bibinfo {title}
  {Observation of a many-body dynamical phase transition with a 53-qubit
  quantum simulator},}\ }\href {https://arxiv.org/abs/1708.01044} {\bibfield
  {journal} {\bibinfo  {journal} {arXiv:1708.01044}\ } (\bibinfo {year}
  {2017})}\BibitemShut {NoStop}%
\end{thebibliography}%

\clearpage
\setcounter{figure}{0}
\makeatletter
\renewcommand{\thefigure}{S\@arabic\c@figure}
\setcounter{equation}{0} \makeatletter
\renewcommand \theequation{S\@arabic\c@equation}
\renewcommand \thetable{S\@arabic\c@table}

\begin{center} 
{\large \bf Supplementary Material for: Machine Learning Bell Nonlocality in
Quantum Many-body Systems}
\end{center}

\section{Bell inequalities with short-range two-body correlators}

In the main text, we have shown that one can use the RBM-based reinforcement
learning to obtain the quantum violations of Ineq. (2). Here, we give
more details on the structure of the neural networks, the DMRG calculations,
and the comparison between obtained results from different methods. 

As discussed in the main text, with properly chosen measurement settings,
the Bell operator corresponding to $\mathcal{I}_{1}$ reduces to the
XXZ-type Hamiltonian $H$ \cite{Tura2017Energy}. With the open boundary
condition, $H$ does not have other obvious symmetry, except that
the total $\Sigma^{z}=\sum\sigma_{k}^{z}$ is conserved. Thus, it
is straightforward to choose a RBM without any symmetry. In this case,
the number of variational parameters is $N+M+N\times M$, which is
large when $N\approx100$ ($\sim10^{4}$). Training this RBM is both
time and memory consuming. For this particular example, we find that
one can instead use a short-range RBM to reduce the number of parameters,
and the accuracy of the final results will not be affected to much.
To be more concrete, we consider a RBM with $M=\alpha N$ ($\alpha$
denotes the hidden unit density and we choose it to be an integer
number for simplicity) and we rearrange the positions of the hidden
neurons, such that at each site there are $\alpha$ hidden neurons
coupling only locally (within range $R$) to the visible neurons.
This significantly reduces the number of parameters, from $O(N^{2})$
to $O(N)$. 

\begin{figure}
\includegraphics[width=0.486\textwidth]{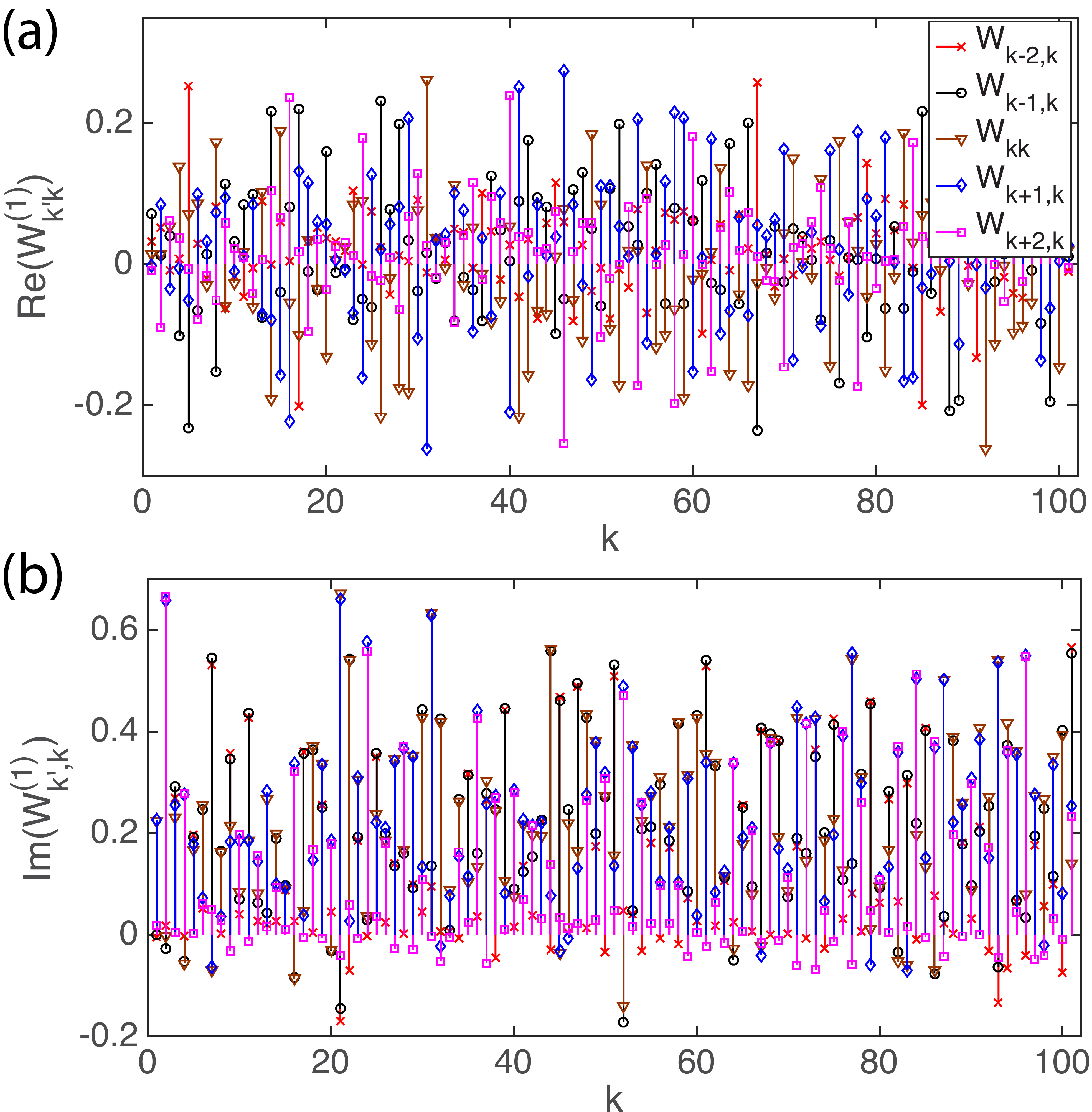}

\caption{The learned weight parameters for representing the ground state of
$H$ with a short-range restricted Boltzmann machine (here, we have
fixed $\alpha=4$ and $R=2$). We denote by $W_{k'k}^{(\eta)}$ ($\eta=1,2,\cdots,\alpha$,
$k',\;k=1,2,\cdots,N$) the coupling strength between the $\eta$-th
hidden neuron at site $k'$ and the visible neuron at sit $k$. (a)
and (b) show the real and imaginary parts of $W_{k'k}^{(1)}$, respectively. They share the same legend.
The parameters specifying $H$ are chosen the same as in Fig. 2(b)
in the main text. \label{fig:The-learned-weight}}
\end{figure}

We begin with a random short-range RBM (i.e., all the internal parameters
are chosen randomly and independently), which typically does not violate
the Ineq. (1) in the main text. We then use a reinforcement learning
algorithm introduced in Ref. \cite{Carleo2016Solving} to optimize
the internal parameters of the RBM. The details of this algorithm
can be found in the Supplementary Materials of Ref. \cite{Carleo2016Solving}.
In Fig. \ref{fig:The-learned-weight}, we plot partial of the weight
parameters for the final trained RBM in Fig. 2(b) in the main text.
Here, only the wight parameters corresponding to the neurons in the
first hidden layer are plotted. The parameters associated to other
hidden layers looks similar and thus are omitted for the sake of conciseness.

In order to characterize the accuracy of the trained RBM, one can
introduced a quantity called the relative error defined as $\epsilon_{\text{rel}}=|E_{0}^{(\text{RBM})}-E_{0}|/E_{0}$,
where $E_{0}$ is the true ground-state energy of $H$ and $E_{0}^{(\text{RBM})}$
denotes the value calculated via the RBM approach. For small system
sizes ($N\leq20$), we find that $\epsilon_{\text{rel}}\sim10^{-4}$
in our calculations. For larger system sizes, we compare our RBM results
with that from DMRG. In our DMRG calculations, we use a MPS representation
of the quantum many-body states and variationally optimizes the MPS
to minimize the ground state energy (see \cite{Schollwock2011Density}
for details). The maximal bond dimension (where we truncate the MPS)
is chosen to be $\chi_{\text{max}}=100$ and we have checked that
the neglected weight for all the truncations are smaller than $10^{-6}$.
We have also examined that the typical variances $\sigma^{2}=\langle H^{2}\rangle-\langle H\rangle^{2}$
is smaller than $10^{-8}$, verifying that the obtained MPS is indeed
an eigenstate of $H$ (up to a negligible error rate). In comparing
our RBM and DMRG results, we find that the relative error $\epsilon_{\text{rel}}=|E_{0}^{(\text{RBM})}-E_{0}^{(\text{DMRG})}|/E_{0}^{(\text{DMRG)}}\sim10^{-3}$.
We mention that the accuracy of the RBM results can be systematically
improved by increasing $R$ and $\alpha$, or the number of iterations
in the training process. In this work of detecting many-body nonlocality
via RBM, high accuracy is not a major concern, hence $\epsilon\sim10^{-3}$
is already sufficient for our purpose. 

\begin{figure}
\includegraphics[width=0.46\textwidth]{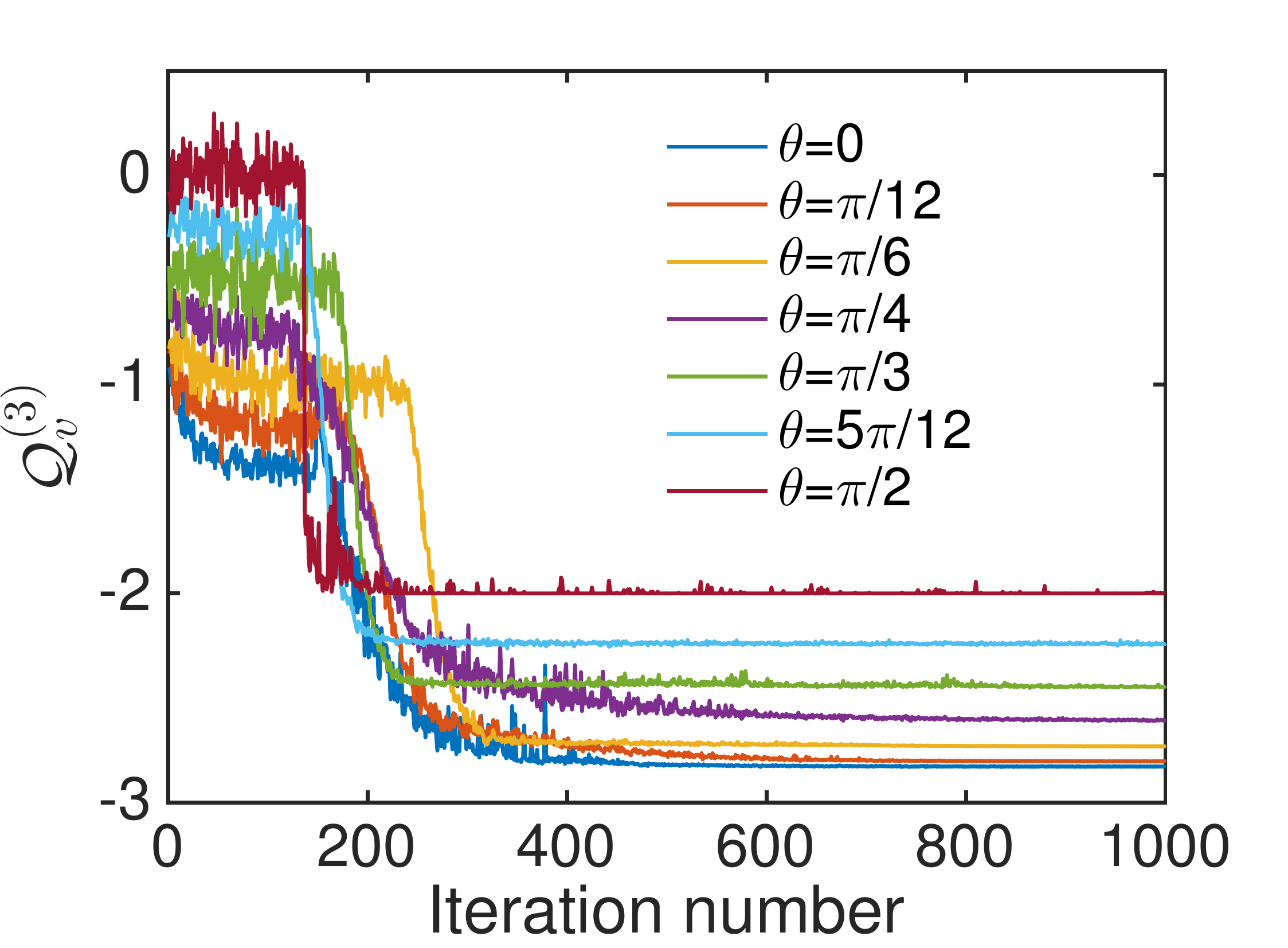}

\caption{RBM-learned quantum expectation value $\mathcal{Q}_{v}^{(3)}$ as
a function of the iteration number, for different measurement angle
$\theta$. \label{fig:QvvsTheta}}
\end{figure}

\section{Bell inequalities with all-to-all two-body correlators}

The Bell inequality in Ineq. (3) in the main text has a permutation
symmetry by construction \cite{Tura2014Detecting}. Thus, its corresponding
Bell operator will also has a permutation symmetry if we choose the
same measurement settings for each party. In this case, it is natural
to use a permutation-invariant RBM to calculate the quantum violations.
This greatly reduces the number of the variational parameters and
the same quantum violations as given in Ref. \cite{Tura2014Detecting}
can be readily obtained. What might be more interesting is the case
in which different party choose different measurement settings. In
this case, the permutation symmetry is violated and it is very challenging
to compute the quantum violations of Ineq. (3). In the main text,
we have considered a scenario where the measurement setting for each
party is random. More specifically, we have chosen the measurement
settings to be: $\mathcal{M}_{0}^{(k)}=\sigma^{z}$ and $\mathcal{M}_{1}^{(k)}=\cos\theta_{k}\sigma^{z}+\sin\theta_{k}\sigma^{x}$
with $\theta_{k}$ being random rotation angles drawn independently
from a uniform distribution $[\theta-\varepsilon,\theta+\varepsilon]$.
Since there is no obvious symmetry for the corresponding Bell operator
and the correlators are all-to-all, we choose the most general RBM
with each hidden neuron connected to all the visible ones.  In plotting
Fig. 3(a) in the main text, we have fixed $\theta=2\pi/3$ and $\varepsilon=0.1$. 

\begin{figure}
\includegraphics[width=0.46\textwidth]{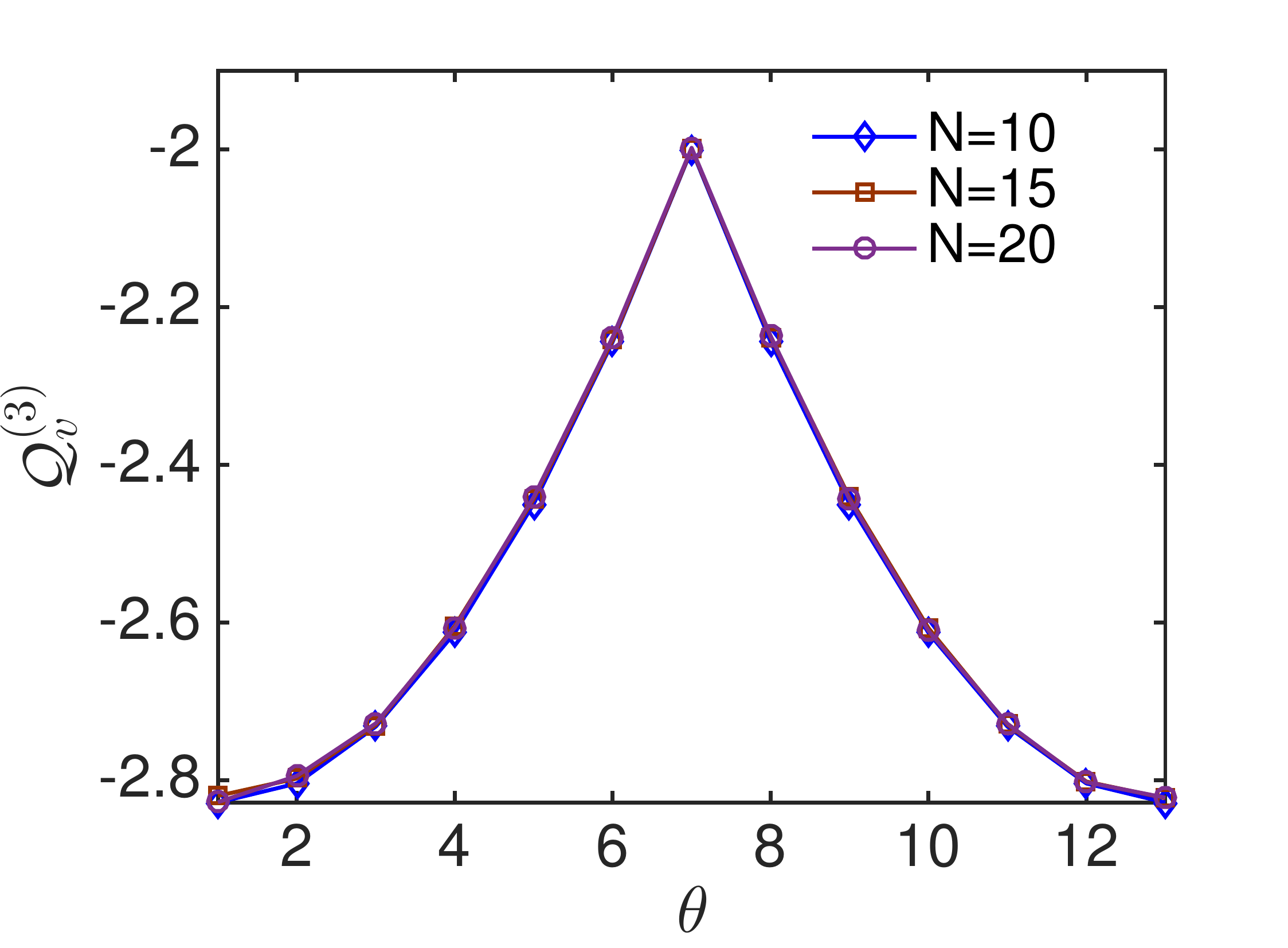}

\caption{RBM-learned quantum violations $\mathcal{Q}_{v}^{(3)}$ as a function
of $\theta$, for different system sizes. \label{fig:QvvsN}}
\end{figure}

\section{\textup{Bell inequalities with multipartite correlators}}

For the Bell inequality (4) in the main text, it is easy to observe
that there is a permutation symmetry between parties indexed from
$2$ to $N$. In addition, the considered measurement settings also
have this symmetry, and thus so does the corresponding Bell operator.
Taking this into consideration, we choose a RBM with the same symmetry:
the RBM contains $M$ hidden neurons and each of them connects to
all visible neurons, but the baise and weight parameters satisfy $a_{2}=a_{3}=\cdots=a_{N}$
and $W_{k',1}=W_{k',2}=\cdots=W_{k',N}$ (for $k'=2,3,\cdots,M$),
respectively. This reduces the number of parameters from $O(MN)$
to $O(M)$ and significantly simplified the calculations. 

In Fig. \ref{fig:QvvsTheta}, we show the RBM-learned $\mathcal{Q}_{v}^{(3)}$
as a function of the iteration number for different measurement angle
$\theta$. It is clear from this figure that $\mathcal{Q}_{v}^{(3)}$
converges rapidly to the corresponding exact values for different
$\theta$. We note that in our calculations we have chosen the learning
rate to be an exponential decaying function of the iteration number,
following Ref. \cite{Carleo2016Solving}. Thus, at the beginning of
the learning process, the learning rate is large. This explains the
large fluctuations at the beginning of the learning process. As the
iteration number increases, the learning rate becomes small and the
curves become smooth. 

In Fig. \ref{fig:QvvsN}, we plot the RBM-learned quantum violations
$\mathcal{Q}_{v}^{(3)}$ as a function of the measurement angle $\theta$
for different system sizes. We find that the quantum violations are
independent of $N$ (up to negligible numerical errors).

\end{document}